\documentclass[aps,prl,twocolumn,groupedaddress,superscriptaddress,nofootinbib,notitlepage,floatfix]{revtex4-2}
\usepackage{times,bm,bbm,bbold,amssymb,amsmath,amsfonts,dsfont,graphics,graphicx,braket,color,xcolor,hyperref}

\hypersetup{colorlinks,linkcolor={blue},citecolor={blue},urlcolor= {blue}}
\newcommand{\ketbra}[2]{|#1\rangle\langle #2|}

\DeclareFontFamily{OT1}{pzc}{}
\DeclareFontShape{OT1}{pzc}{m}{it}
              {<-> s * [1.25] pzcmi7t}{}
\DeclareMathAlphabet{\mathpzc}{OT1}{pzc}
                                 {m}{it}

\newtheorem{lemma}{Lemma}
\newcommand{\ignore}[1]{}
\usepackage[T1]{fontenc}
\usepackage{newtxtext,newtxmath}

\begin{document}

\title{Temperature in Nonequilibrium Quantum Systems}

\author{S. Alipour}
\email{sahar.alipour@aalto.fi}
\affiliation{QTF Center of Excellence, Department of Applied Physics, Aalto University, FI-00076 Aalto, Finland}

\author{F. Benatti}
\affiliation{Department of Physics, University of Trieste, I-34151 Trieste, Italy}
\affiliation{National Institute for Nuclear Physics (INFN), Trieste Section, I-34151 Trieste, Italy}

\author{M. Afsary}
\affiliation{Department of Physics, Sharif University of Technology, Tehran 14588, Iran}

\author{F. Bakhshinezhad}
\affiliation{Department of Physics and NanoLund, Lund University, Box 188, SE-221 00, Lund, Sweden}

\author{M. Ramezani}
\affiliation{Department of Physics, Sharif University of Technology, Tehran 14588, Iran}

\author{T. Ala-Nissila}
\affiliation{QTF Center of Excellence, Department of Applied Physics, Aalto University, FI-00076 Aalto, Finland}
\affiliation{Interdisciplinary Centre for Mathematical Modelling and Department of Mathematical Sciences, Loughborough University, Loughborough, Leicestershire LE11 3TU, UK}

\author{A. T. Rezakhani}
\affiliation{Department of Physics, Sharif University of Technology, Tehran 14588, Iran}
\affiliation{School of Physics, Institute for Research in Fundamental Sciences (IPM), Tehran 19538, Iran}

\begin{abstract}
\noindent We extend on ideas from standard thermodynamics to show that temperature can be assigned to a general nonequilibrium quantum system. By choosing a physically motivated complete set of observables and expanding the system state thereupon, one can read a set of relevant, independent thermodynamic variables which include internal energy. This expansion allows us to read a nonequilibrium temperature as the partial derivative of the von Neumann entropy with respect to internal energy. We show that this definition of temperature is one of a set of thermodynamics parameters unambiguously describing the system state. It has appealing features such as positivity for passive states and consistency with the standard temperature for thermal states. By attributing temperature to correlations in a bipartite system, we obtain a universal relation which connects the temperatures of subsystems, total system as a whole, and correlation. All these temperatures can be different even when the composite system is in a well-defined Gibbsian thermal state. 
\end{abstract}
\date{\today}
\maketitle

\textit{Introduction.---}Circuit quantum thermodynamics is an important aspect of future superconducting quantum computers, where controlled cooling and heating quantum devices in the range of millikelvins is necessary \cite{Pekola-thermo-circuit}. Recent progress in quantum technologies with the advent of novel quantum devices is now in the stage that measuring heat generated from absorption of single photons is becoming possible, and accurate calorimeters and bolometers working in the ultimate energy resolution regime are now available \cite{Pekola-ultimate-resol, Mottonen-Nature-bolometer}. In addition to such technological developments, there have been numerous studies attempting to extend standard thermodynamics to the quantum regime and to understand how quantum features may affect thermodynamic laws or properties \cite{gemmer2009quantum, book:Malher, book:Qthermo, book:Deffner-Campbell, Allahverdyan--, Goldstein, Goold16, Gogolin-Eisert, Allahverdyan-, Scully-SingleHeatBath, SciRep, Vinjanampathy-Modi-initialCorr, bera-GeneralizedLawsThermo, Su-heatCoherence, Bakhshinezhad_2019, Manzano-work-corr, Lutz-Reverse-Heat-flow, unreliability-mutual-info, Anders-coherence, Paternostro-work-initial-coherence, SciRep-entropy-production, Strasberg-entroproduct, Manzano-martingale-entropyProduction, Esposito09, Campisi11, Funo18, ATR-qft2, delCampo-fluctuation-battery, Anders-Lutz-Landauer, Kosloff13, delcampo18, Rivas-nonEquilibMeanForce, ATR-carnot}. 

Despite progress in quantum thermodynamics, there still exist several incompatible definitions of basic concepts such as heat, work, entropy, and temperature \cite{Alicki, Spohn-EP, Kurizki-ergo, Polkovnikov-diag-entr, entropic-based-heat-work, Ahmadi-refined, safranek, Sampaio2018}. Among these concepts, defining temperature is specially challenging and despite studies on how to measure temperature in quantum systems \cite{mehboudi2019thermometry, Correa-manybody-thermometry, Bakhshinezhad-cooling, PRXCotler-virtual-cooling, Raeisi-algorCooling}, there is still no full understanding on how the concept of temperature can be extended to the quantum regime. A major difficulty here is that open quantum systems are often in nonequilibrium states, and strongly coupled and correlated with their environments \cite{ULL, ULL-2, strasberg2016nonequilibrium, alipour2020shortcuts}. Standard thermodynamics often ignores coupling and correlations between the system and its environment or bath (see also recent studies addressing this issue \cite{SciRep,Rivas-nonEquilibMeanForce, Vinjanampathy-Modi-initialCorr, Seifert-FirstSecondLawStrongCoupling, Jarzynski-StronglyCoupledThermo, bera-GeneralizedLawsThermo, hsiang-StrongCoupling-OperatorThermodynamics, Perarnau-Llobet-StrongCouplingThermo, Talkner-StrongCouplingThermo-Classic-Quantum, HosseinNejad-OlayaCastro, Strasberg-StrongCoupling-nonMarkovian}). In general, coupling with the environment leads to deviation of the asymptotic state of the system from the Gibbsian form \cite{nazir, kawai}. This leads to the issues of locality or nonintensivity of temperature, which makes defining subsystem temperature nontrivial even when the total system is in an equilibrium state and has a well-defined temperature \cite{Acin-intensiveT, Gogolin-LocalityTemp, LocalityTempSpinChain, Vaezi-localT}. The situation appears to be particularly challenging when the system is not in equilibrium \cite{Kurchan, puglisi2017temperature, Martens-FlucDissipTemp}. 

There is a well-motivated definition of temperature in classical thermodynamics based on the partial derivative of the entropy $S$ with respect to the internal energy $U$ \cite{book:Callen},
\begin{align}
\frac{1}{T}&=\left(\frac{\partial S}{\partial U}\right)_{N,V,\ldots},
\label{classical-T}
\end{align}
where $N$ is the number of particles and $V$ is the volume. Despite attempts to generalize the definition \eqref{classical-T} to quantum systems \cite{vallejo2020temperature}, it has remained elusive due to the ambiguity in defining $N$ and $V$. In practice, however, temperature in nonequilibrium quantum systems has been mostly defined based on comparing the state of the system with a suitable reference Gibbs state. For example, a temperature is assigned to a system in a non-Gibbsian state by matching its average energy with the average energy of the same system in a Gibbs state \cite{Scully-SingleHeatBath, book:Fick-Sauermann}. 

In the following, we argue that temperature can still be defined in quantum systems using the same definition as in Eq. \eqref{classical-T}. By considering the density matrix of the system as the basis of any thermodynamic state function of the system,  and expanding the density matrix in terms of a set of independent operators, we introduce a set of independent variables for a given quantum system. This makes the calculation of the partial derivative feasible and a closed relation for temperature in general quantum systems can be obtained. 

\textit{Temperature in quantum systems.---}To formulate a notion of temperature that is also applicable in nonequilibrium settings, we start by representing the system density matrix $\varrho$ in terms of a set of independent parameters associated with pre-assigned relevant properties of the quantum system. For a $d$-level system, this expansion can be done by means of a set of traceless observables $\{O^{(i)}\}_{i=1}^{d^{2}-1}$ such that $\mathrm{Tr}[O^{{(i)}\dag} O^{(j)}]=\delta_{ij}$, augmented by the normalized identity $O^{(0)}=\mathbbmss{I}/\sqrt{d}$. These operators constitute a Hilbert-Schmidt orthonormal basis in the algebra of $d \times d$ operators such that one can expand any operator thereupon; e.g., $\varrho=\sum_{i=0}^{d^{2}-1} x_{i}\,O^{(i)}$. The mean values $x_{i}=\mathrm{Tr}[\varrho\,O^{(i)}]$ with $i\geqslant 1$ are the relevant parameters whose variations characterize the reaction of the system to external changes. For example, in a thermodynamic context, an important property is the internal energy, so we shall take $O^{(1)}$ as the traceless part of the system Hamiltonian $H$ such that   
\begin{equation}
O^{(1)}=(1/h) \big(H-\mathrm{Tr}[H]\mathbbmss{I}/d\big),
\label{tracelessH}
\end{equation}
with $h:=\sqrt{\mathrm{Tr}[H^{2}]-\mathrm{Tr}[H]^{2}/d}$. Note that when the system interacts with an environment, the relevant Hamiltonian  $H$ of the system is the effective Hamiltonian \cite{SciRep} rather than the bare Hamiltonian $H_{0}$ of the unperturbed system. For example, for an open system with Markovian dynamics, $H=H_{0}+H_{\mathrm{Lamb} }$, where $H_{\mathrm{Lamb}}$ is the environment-induced Lamb shift \cite{BreuerBook}. Interestingly, as we show in Ref. \cite{SM}, despite the freedom in choosing $O^{(2)},\ldots, O^{(d^{2}-1)}$, our formalism is independent of this choice. Note also that $H$ and the other observables $O^{(i>1)}$ may be time-dependent, hence all relations we obtain in the sequel are also time-dependent. However, to simplify the notation we omit time dependence almost everywhere unless needed to avoid ambiguity.

Expanding $\varrho$ in terms of the observable basis $\{O^{(i)}\}$ gives $\varrho(\{ x_{i}\}) =\textstyle{\sum_{i=0}^{d^{2}-1}}  x_{i}\, O^{(i)}$. The von Neumann entropy $S$ and the internal energy $U$ can also be rewritten in terms of $\{x_{i}\}$ as
\begin{gather}
 S=-\mathrm{Tr}[\varrho \log \varrho] = - \textstyle{\sum_{i}} \mathrm{Tr}[O^{(i)} \log \varrho]\, x_{i}. 
\label{sa} \\
U= \mathrm{Tr}[\varrho H]= \textstyle{\sum_{i>0}}  \mathrm{Tr}[O^{(i)}\,H]\, x_{i} = h\, x_{1}+\mathrm{Tr}[H]/d.
\label{U-cov-x1}
\end{gather}
This relation shows that internal energy (modified by the prefactor $1/h$) can particularly be identified with the first independent variable $x_{1}$. Thus, the entropy can be recast as
\begin{equation}
S=-\frac{1}{h}\mathrm{Tr}[O^{(1)} \log\varrho]\, U - \textstyle{\sum_{i\geqslant 2}} \mathrm{Tr}[O^{(i)}\log\varrho]\, x_{i}+\mathcal{I},
\label{s-new}
\end{equation}
where the last term $\mathcal{I}$ denotes variable-independent, irrelevant terms. Thus, we can obtain the nonequilibrium temperature as
\begin{align}
\frac{1}{T} &=\left(\frac{\partial S}{\partial U}\right)_{x_{2}, x_{3},\ldots}= -\frac{1}{h}\mathrm{Tr}[O^{(1)} \log\varrho]
\label{temp-new}\\
&= \frac{\mathrm{Cov}(H,\mathbbmss{H})}{\Delta^{2} H}\ ,
\label{def:T}
\end{align}
where $\mathrm{Cov}(X, Y)=\mathrm{Tr}[(\mathbbmss{I}/d)X Y]\,-\,\mathrm{Tr}[(\mathbbmss{I}/d)X]\,\mathrm{Tr}[(\mathbbmss{I}/d)Y]$ and
$\Delta^{2} H =\mathrm{Tr}[(\mathbbmss{I}/d)H^{2}]\,-\,\left(\mathrm{Tr}[(\mathbbmss{I}/d)H]\right)^{2}$ are covariance and mean-square value evaluated with respect to the maximally mixed state $\mathbbmss{I}/d$, and $\mathbbmss{H}=-\log \varrho$. Note that when $\varrho$ is the reduced density matrix of a bipartite \textit{pure} state, $\mathbbmss{H}$ is known as the \textit{entanglement Hamiltonian}, whose spectrum can be used to reveal quantum phase transitions \cite{ent-ham, ent-spec}. 

\textit{Properties of temperature.---}(i) Consistency with the temperature in Gibbsian states: if the system is in a thermal state given by $\varrho^{\beta}=(1/Z)e^{-\beta H}$, it is straightforward to see from Eq. \eqref{def:T} that $T=1/\beta$ (assuming $k_{B} \equiv 1$), which perfectly coincides with our expectation of the temperature in such systems. (ii) Vanishing (diverging) temperature for pure (maximally mixed) states: It is straightforward from Eq. (\ref{def:T}) that, regardless of the system Hamiltonian, for pure states $|\psi\rangle$ we have $T=0$ and for maximally mixed states $\mathbbmss{I}/d$ we obtain $T=\infty$. (ii) Positivity of $T$ for passive states: consider a passive quantum state $\varrho =\sum_{i=1}^{d} \eta_{i} |e_{i} \rangle \langle e_{i}|$ where $\eta_{i+1}\leqslant \eta_{i}$, such that its Hamiltonian reads as $H=\sum_{i=1}^{d} E_{i} |e_{i} \rangle \langle e_{i}|$ where $E_{i+1}\geqslant E_{i}$. To show that $T$ is positive, we need to prove that $\mathrm{Cov}(H,-\log \varrho)\geqslant 0$. The latter hinges on having $\mathrm{Tr}[AB]\geqslant 0$ for any traceless Hermitian $A$ and positive $B$ \cite{SM}; simply choose $A=(H-\mathrm{Tr}[H]\mathbbmss{I})/d$ and $B= -\log \varrho$. Note that the temperature for a general nonpassive state is not necessarily positive. As in standard thermodynamics \cite{puglisi2017temperature, Smorodinsky-temperature, Purcell-NeqTemp, Oja-NegTemp, Medley-NegTemp, braun-negativeTemp}, negative temperatures are physically possible. (iv) Invariance with respect to trivial extension: if we trivially extend the Hilbert space of the system, the assigned temperature remains unchanged. This can be seen by changing $H \to H\otimes \mathbbmss{I}$ and normalizing the corresponding $O^{(1)} \otimes \mathbbmss{I}$ operator in Eq. \eqref{temp-new}. (v) Consistency with the definition of temperature obtained from the extended Gibbs state: Any full-rank quantum state can be written as $\varrho=e^{\log \varrho}$. By expanding $\log \varrho$ in terms of the Hamiltonian and the observables $\{O^{(i)}\}$ orthogonal to $O^{(1)}$ as $\log \varrho=-(\log Z) \mathbbmss{I} -\beta H +\sum_{i>1} c_{i} O^{(i)}$, it is seen that $\beta=1/T$ and $Z=e^{-\mathrm{Tr}[\log \varrho]}$, so that $\varrho=(1/Z)e^{-\beta H +\sum_{i>1} c_{i} O^{(i)}}$ looks like a generalized Gibbs state. Note that, however, unlike standard generalized Gibbs states \cite{Schmiedmayer-GeneralizedGibbsObservation} which are usually introduced in relation to stationary states and wherein the observables $O^{(i)}$ are constants of motion, here the generalized Gibbs from depends on time and the observables $O^{(i)}$ are basis operators, which need not commute with the Hamiltonian.

\textit{Example: Two interacting qubits in thermal equilibrium.}---Consider two interacting qubits $\mathsf{S}$ and $\mathsf{B}$ with  Hamiltonians $H_{\mathsf{S}}=(\omega_{\mathsf{S}}/2)\sigma_{z}$ and $H_{\mathsf{B}}=(\omega_{\mathsf{B}}/2) \sigma_{z}$ that interact via $H_{\mathrm{I}}=\lambda( \sigma_{+} \otimes \sigma_{-} +\sigma_{-} \otimes \sigma_{+})$, where $\sigma_{\pm}=\sigma_x\pm i\sigma_y$, with $\sigma_{x}$, $\sigma_{y}$, and $\sigma_{z}$ being the Pauli operators, and we assume $\omega_{\mathsf{S}}\geqslant \omega_{\mathsf{B}}\geqslant 0$, and $\lambda>0$ is a constant. We assume that the total system is in thermal equilibrium at temperature $T=1/\beta$, 
\begin{equation}
\label{Gibbs}
\varrho_{\mathsf{SB}}^\beta=(1/Z ) e^{-\beta H_{\mathsf{SB}}},
\end{equation}
where $Z = \mathrm{Tr}[e^{-\beta_{\mathsf{SB}} H_{\mathsf{SB}}}] = 2[\cosh(\beta\Delta_{+}) + \cosh(\beta\eta)]$, with $\Delta_{\pm}=(\omega_{\mathsf{S}} \pm\omega_{\mathsf{B}})/2$ and $\eta=\sqrt{\Delta_{-}^{2}+16\lambda^{2}}$. One can in general associate with the qubits the effective Hamiltonians $H_{\mathsf{S}}^{(\mathrm{eff})}=H_{\mathsf{S}} + \mathrm{Tr}[\varrho_{\mathsf{B}}\,H_{\mathrm{I}}]$ and similarly for $H_{\mathsf{B}}^{(\mathrm{eff})}$ \cite{SciRep}, where here reduce to $H_{\mathsf{S}}^{(\mathrm{eff})}=H_{\mathsf{S}}$ and $H_{\mathsf{B}}^{(\mathrm{eff})}=H_{\mathsf{B}}$. In this example, the states of the qubits are also Gibbsian, 
\begin{align}
\label{2qubits3a}
\varrho_{\mathsf{S}}&=\mathrm{Tr}_{\mathsf{B}}[\varrho_{\mathsf{SB}}^\beta]=(1/Z_{\mathsf{S}} )  e^{-\beta_{\mathsf{S}} H_{\mathsf{S}}}\\ 
\label{2qubits3a-2}
\varrho_{\mathsf{B}}&=\mathrm{Tr}_{\mathsf{S}}[\varrho_{\mathsf{SB}}^\beta]=(1/Z_{\mathsf{B}} )  e^{-\beta_{\mathsf{B}} H_{\mathsf{B}}},
\end{align}
where  $\beta_{\mathsf{S}}= (1/\omega_{\mathsf{S}}) \log(\mu^{(1)}_{+}/\mu^{(1)}_{-})$ and $\beta_{\mathsf{B}}= (1/\omega_{\mathsf{B}}) \log(\mu^{(2)}_{-}/\mu^{(2)}_{+})$, with $\mu^{(1)}_{\pm}$ and $\mu^{(2)}_{\pm}$, respectively, being the eigenvalues of $\varrho_{\mathsf{S}}$ and $\varrho_{\mathsf{B}}$, 
\begin{align}
\label{2qubits4}
\mu^{(1)}_{\pm}&=(1/Z )[e^{\pm\beta\Delta_{+}} + \cosh(\beta\eta) \pm (\Delta_{-}/\eta) \sinh(\beta\eta)],\\
\label{2qubits4a}
\mu^{(2)}_{\pm}&=(1/Z )[ e^{\mp\beta\Delta_{+}} + \cosh (\beta\eta) \pm (\Delta_{-}/\eta) \sinh(\beta\eta)].
\end{align}
Thus, from Eq. \eqref{def:T}, the system and bath temperatures are obtained as $T_{\mathsf{S}}=\beta_{\mathsf{S}}^{-1}$ and $T_{\mathsf{B}}=\beta_{\mathsf{B}}^{-1}$, as expected.

Since $\mu^{(1)}_{+}\geqslant \mu^{(1)}_{-}$, we have $T_{\mathsf{S}}\geqslant 0$. In contrast, the sign of $T_{\mathsf{B}}$ depends on $\lambda$ and $\Delta_{\pm}$. Such an asymmetry between the two temperatures follows from our initial choice $\omega_{\mathsf{S}}\geqslant \omega_{\mathsf{B}}$, and it disappears when the two level spacings are the same, in which case $\Delta_{-}=0$ and $\mu^{(1)}_{+} / \mu^{(1)}_{-} = \mu^{(2)}_{-} / \mu^{(2)}_{+} \geqslant 1$. The two temperatures are also equal when the interaction is switched off; then $\mu^{(1)}_{+}/\mu^{(1)}_{-}= e^{\beta \omega_{\mathsf{S}}}$ and $\mu^{(2)}_{-}/\mu^{(2)}_{+}=e^{\beta\omega_{\mathsf{B}}}$ and hence---as expected---the two temperatures coincide with $T=1/\beta$.

\textit{Temperature of correlation.}---As we have observed in the previous example, in general, there can be a discrepancy between the temperatures of the different subsystems of a composite system and also between the temperature of the subsystems and the temperature of the total system itself. This feature may persist even when the composite system is in a thermal state and the subsystems are the same and again in thermal states. This discrepancy begs further exploration. 

It is known that in a composite system, a part of the total internal energy given by 
\begin{align}
U_{\chi}=\mathrm{Tr}[\chi\,H^{(\mathrm{eff})}_{\mathrm{I}}]
\label{U-chi}
\end{align}
 is inaccessible to the local subsystems, which hence should be assigned to the correlation $\chi= \varrho_{\mathsf{SB}}-\varrho_{\mathsf{S}}\otimes\varrho_{\mathsf{B}}$ between the subsystems \cite{SciRep, SciRep-entropy-production}. Here $H^{(\mathrm{eff})}_{\mathrm{I}}$ represents an effective interaction Hamiltonian $H^{(\mathrm{eff})}_{\mathrm{I}}=H_{\mathrm{I}}-\mathrm{Tr}_{\mathsf{S}}[\varrho_{\mathsf{S}}\,H_{\mathrm{I}}]-\mathrm{Tr}_{\mathsf{B}}[\varrho_{\mathsf{B}}\,H_{\mathrm{I}}] + \mathrm{Tr}[\varrho_{\mathsf{S}} \otimes \varrho_{\mathsf{B}}\, H_{\mathrm{I}}] \mathbbmss{I}_{\mathsf{SB}}$, that is, the interaction Hamiltonian $H_{\mathrm{I}}$ is modified by subtracting the Lamb-shift-like Hamiltonians and adding a scalar quantity, such that $\mathrm{Tr}_{\mathsf{S}}[\varrho_{\mathsf{S}}\,H^{(\mathrm{eff})}_{\mathrm{I}}] = \mathrm{Tr}_{\mathsf{B}}[\varrho_{\mathsf{B}} \,H^{(\mathrm{eff})}_{\mathrm{I}}]=0$. We note that $\mathrm{Tr}[\chi H^{(\mathrm{eff})}_{\mathrm{I}}] \equiv \mathrm{Tr}[\chi H_{\mathrm{I}}]$ and that $\mathrm{Tr}[\chi H_{\mathrm{I}}] = \mathrm{Tr}[\varrho_{\mathsf{SB}} H_{\mathsf{SB}}]-\mathrm{Tr}[\varrho_{\mathsf{S}}\otimes \varrho_{\mathsf{B}}  H_{\mathsf{SB}}]$. This also justifies that $U_{\chi}$, the difference between the energy of the correlated state $\varrho_{\mathsf{SB}}$ and the uncorrelated state $\varrho_{\mathsf{S}} \otimes \varrho_{\mathsf{B}}$, should naturally be assigned to correlations and has the role of a binding energy. It has been recently shown that $\chi$ can be explicitly unraveled from master equations for the dynamics of the subsystem \cite{ULL-2}. 

Correlations between the subsystems can be characterized by mutual information $S_{\chi}= S(\varrho_{\mathsf{SB}} \Vert \varrho_{\mathsf{S}}\otimes \varrho_{\mathsf{B}} )$, where $S(\varrho_{1}\Vert \varrho_{2})=\mathrm {Tr}[\varrho_{1}(\log\varrho_{1}-\log\varrho_{2})]$ is the relative entropy. Since $S_{\chi}$ can also be recast as $S_{\chi}= S(\varrho_{\mathsf{S}} \otimes \varrho_{\mathsf{B}}) - S(\varrho_{\mathsf{SB}})$, it can be considered as the entropy assigned to correlations (for some pertinent caveats, see Ref. \cite{unreliability-mutual-info}).

Having an energy and an entropy assigned to the correlation $\chi$, we can also associate a temperature with it by using the thermodynamics definition \eqref{classical-T}. To do so, we choose a traceless basis operator $O_{\mathrm{I}}$ of the form 
\begin{align}
O_{\mathrm{I}}&= (1/h_{\mathrm{I}}) \big(H^{(\mathrm{eff})}_{\mathrm{I}}-\mathrm{Tr}[H^{(\mathrm{eff})}_{\mathrm{I}}]/d_{\mathsf{SB}}\big),
\end{align}
where $h_{\mathrm{I}} =\sqrt{\mathrm{Tr}[(H^{(\mathrm{eff})}_{\mathrm{I}})^{2}]-\mathrm{Tr}[H^{(\mathrm{eff})}_{\mathrm{I}}]^{2}/d_{\mathsf{SB}}}$ ensures $\mathrm{Tr}[O_{\mathrm{I}}^{2}]=1$. We expand $\varrho_{\mathsf{SB}}$ in terms of a set of independent operators $\{\mathbbmss{I}_{\mathsf{S}}\otimes \mathbbmss{I}_{\mathsf{B}}, O_{\mathsf{S}}\otimes \mathbbmss{I}_{\mathsf{B}}, \mathbbmss{I}_{\mathsf{S}}\otimes O_{\mathsf{B}}, O_{\mathrm{I}},R^{(i)}\}_{i=1}^{d_{\mathsf{SB}}^{2}-4}$, where $O_{\mathsf{S}}$ and $O_{\mathsf{B}}$ are normalized traceless operators constructed from $H^{(\mathrm{eff})}_{\mathsf{S}}$ and $H^{(\mathrm{eff})}_{\mathsf{B}}$, respectively (with the normalization factors $h_{\mathsf{S}}$ and $h_{\mathsf{B}}$), and $\{R^{(i)}\}$ is the set of the remaining orthonormal operators, which are orthogonal to the subspace $\mathpzc{L}_{H}=\{O_{\mathsf{S}}\otimes \mathbbmss{I}_{\mathsf{B}}, \mathbbmss{I}_{\mathsf{S}}\otimes O_{\mathsf{B}}, O_{\mathrm{I}}\}$ and together with it form a complete set of operators for the total system. Using this set we define a set of independent variables $\{U_{\mathsf{S}}, U_{\mathsf{B}}, U_{\chi},r_{i}\}_{i=1}^{d_{\mathsf{SB}}^{2}-4}$ such that 
\begin{align}
&S_{\chi}=-\frac{G}{h_{\chi}^{2}} \mathrm{Tr}[\mathbbmss{H}_{\mathrm{I}} O_{\mathrm{I}}]
- \left(\frac{U_{\mathsf{S}}}{2h_{\mathsf{S}}}- \frac{G\,\mathrm{Tr}[O_{\mathrm{I}} O_{\mathsf{S}} \otimes \mathbbmss{I}_{\mathsf{B}} ]}{d_{\mathsf{B}} h_{\chi}^{2}}\right)\mathrm{Tr}[\mathbbmss{H}_{\mathrm{I}} O_{\mathsf{S}} \otimes \mathbbmss{I}_{\mathsf{B}} ]\nonumber\\
&\,- \left(\frac{U_{\mathsf{B}}}{2h_{\mathsf{B}}}- \frac{G\,\mathrm{Tr}[O_{\mathrm{I}} \mathbbmss{I}_{\mathsf{S}} \otimes O_{\mathsf{B}}]}{d_{\mathsf{S}} h_{\chi}^{2}}\right) \mathrm{Tr}[\mathbbmss{H}_{\mathrm{I}} \mathbbmss{I}_{\mathsf{S}} \otimes O_{\mathsf{B}}] + \textstyle{\sum_{i}} r_{i} \,\mathrm{Tr}[\mathbbmss{H}_{\mathrm{I}} R^{(i)}] \nonumber\\
&\, +\mathcal{I},
\label{Schi-expanded}
\end{align}
where $G(U_{\mathsf{S}}, U_{\mathsf{B}}, U_{\chi})=U_{\chi}-\frac{\mathrm{Tr}[O_{\mathrm{I}} O_{\mathsf{S}} \otimes \mathbbmss{I}_{\mathsf{B}} ]}{d_{\mathsf{B}}h_{\mathsf{S}}}U_{\mathsf{S}}-\frac{\mathrm{Tr}[O_{\mathrm{I}} \mathbbmss{I}_{\mathsf{S}} \otimes O_{\mathsf{B}}]}{d_{\mathsf{S}}h_{\mathsf{B}}}U_{\mathsf{B}}$, $h_{\chi}=(1-\frac{\mathrm{Tr}[O_{\mathrm{I}} \, O_{\mathsf{S}}\otimes \mathbbmss{I}_{\mathsf{B}} ]^{2}}{d_{\mathsf{B}}}-\frac{\mathrm{Tr}[O_{\mathrm{I}} \,\mathbbmss{I}_{\mathsf{S}} \otimes  O_{\mathsf{B}}]^{2}}{d_{\mathsf{S}} })^{1/2}$, $r_{i}=\mathrm{Tr}[\varrho_{\mathsf{SB}} R^{(i)}]$, $\mathbbmss{H}_{\mathrm{I}}=\mathbbmss{H}_{\mathsf{SB}}-\mathbbmss{H}_{\mathsf{S}}\otimes \mathbbmss{I}_{\mathsf{B}}-\mathbbmss{I}_{\mathsf{S}}\otimes \mathbbmss{H}_{\mathsf{B}}$ and $\mathcal{I}$ again denotes irrelevant terms. Equation \eqref{Schi-expanded} gives the correlation temperature as
\begin{align}
\frac{1}{T_{\chi}}=&\left(\frac{\partial S_{\chi}}{\partial U_{\chi}}\right)_{U_{\mathsf{S}},U_{\mathsf{B}},\{r_{i}\}}=-\frac{1}{h_{\chi}^{2}}\mathrm{Tr}[\mathbbmss{H}_{\mathrm{I}} O_{\mathrm{I}}] +\frac{1}{d_{\mathsf{B}} h_{\chi}^{2}} \mathrm{Tr}[O_{\mathrm{I}} O_{\mathsf{S}} \otimes \mathbbmss{I}_{\mathsf{B}}] \nonumber\\
& \times \mathrm{Tr}[O_{\mathsf{S}} \otimes \mathbbmss{I}_{\mathsf{B}} \mathbbmss{H}_{\mathrm{I}}]
+\frac{1}{d_{\mathsf{S}} h_{\chi}^{2}} \mathrm{Tr}[O_{\mathrm{I}} \mathbbmss{I}_{\mathsf{S}} \otimes O_{\mathsf{B}}] \, \mathrm{Tr}[\mathbbmss{I}_{\mathsf{S}} \otimes O_{\mathsf{B}}\mathbbmss{H}_{\mathrm{I}}].
\label{T_chi}
\end{align}

It can be observed that when at any instant of time the correlation between the interacting parts vanishes (i.e, $\log \varrho_{\mathsf{SB}}=\log \varrho_{\mathsf{S}}+\log \varrho_{\mathsf{B}}$ or equivalently $\mathbbmss{H}_{\mathrm{I}}=0$), we have $T_{\chi} \to \infty$. For the opposite case, where $\varrho_{\mathsf{SB}}$ is a maximally entangled pure state, we conclude that $T_{\chi} \to 0$. 

\begin{figure}[tp]
\includegraphics[scale=0.3]{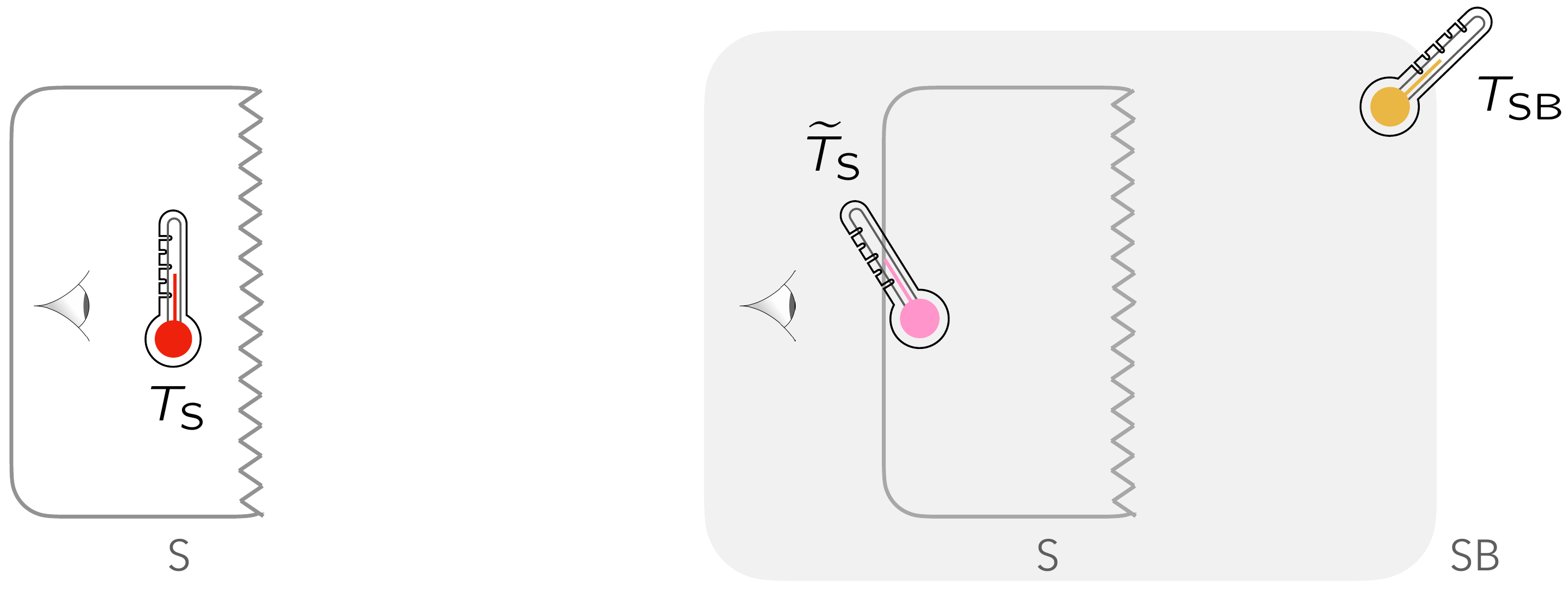}
\caption{Schematic of the different definitions of temperatures $T_{\mathsf{S}}$, $\widetilde{T}_{\mathsf{S}}$, and $T_{\mathsf{SB}}$. See text for details.}
\label{fig:fig-1}
\end{figure}

\textit{Relation between the temperatures of the subsystems, correlation, and the total system}.---To unravel how all the different definitions of temperature here are related, we start from $S_{\mathsf{SB}}=S_{\mathsf{S}}+S_{\mathsf{B}}-S_{\chi}$ and take partial derivative of both sides with respect to $U_{\mathsf{SB}}$, keeping the other independent variables constant. This yields
\begin{align}
\frac{K_{\mathsf{SB}}}{T_{\mathsf{SB}}} = \frac{b_{\mathsf{S}}}{\widetilde{T}_{\mathsf{S}}} + \frac{b_{\mathsf{B}}}{\widetilde{T}_{\mathsf{B}}} - \frac{K_{\chi}}{T_{\chi}},
\label{Ts-connection-main}
\end{align}
where 
\begin{align}
 \frac{1}{\widetilde{T}_{\mathsf{S}}}&:=\left(\frac{\partial S_{\mathsf{SB}}}{\partial U_{\mathsf{S}}}\right)_{U_{\mathsf{B}},U_{\chi},\{r_{i}\}}=\frac{1}{T_{\mathsf{S}}}-\left(\frac{\partial S_{\chi}}{\partial U_{\mathsf{S}}}\right)_{U_{\mathsf{B}},U_{\chi},\{r_{i}\}},
 \\
 \frac{1}{\widetilde{T}_{\mathsf{B}}}&:=\left(\frac{\partial S_{\mathsf{SB}}}{\partial U_{\mathsf{B}}}\right)_{U_{\mathsf{S}},U_{\chi},\{r_{i}\}}=\frac{1}{T_{\mathsf{B}}}-\left(\frac{\partial S_{\chi}}{\partial U_{\mathsf{B}}}\right)_{U_{\mathsf{S}},U_{\chi},\{r_{i}\}}, 
\end{align}
and $K_{\mathsf{SB}}$, $K_{\chi}$, $b_{\mathsf{S}}$, and $b_{\mathsf{B}}$ are appropriate coefficients which depend only on the local and interaction Hamiltonians and the dimensions of the subsystems; for the explicit forms see Ref.~\cite{SM}. Here $\widetilde{T}_{\mathsf{S}}$ and $\widetilde{T}_{\mathsf{B}}$ can be considered as the temperatures assigned to the subsystems when we have access to the total information $S_{\mathsf{SB}}$. In contrast, $T_{\mathsf{S}}$ and $T_{\mathsf{B}}$ are the temperatures when only the local information $S_{\mathsf{S}}$ and $S_{\mathsf{B}}$ are accessible (cf. Fig. \ref{fig:fig-1} for a schematic).

\textit{Example (cntd).}---Inserting Eqs. \eqref{Gibbs} -- \eqref{2qubits3a-2} into Eq. \eqref{T_chi}, we obtain $1/T_{\chi}=-\beta$ and $1/\widetilde{T}_{\mathsf{S}}=1/\widetilde{T}_{\mathsf{B}}=\beta$. Noting that $h_{\mathsf{SB}}=(\omega_{\mathsf{S}}^{2}+\omega_{\mathsf{B}}^{2}+2\lambda^{2})^{1/2}$, $h_{\mathsf{S}}=\omega_{\mathsf{S}}/\sqrt{2}$, $h_{\mathsf{B}}=\omega_{\mathsf{B}}/\sqrt{2}$, $h_{\mathrm{I}}=\sqrt{2}\lambda$, and $h_{\chi}=1$, we obtain $b_{\mathsf{S}}=\omega_{\mathsf{S}}^{2}/\mathsf{D}$, $b_{\mathsf{B}}=\omega_{\mathsf{B}}^{2}/\mathsf{D}$, $K_{\mathsf{SB}}=(\omega_{\mathsf{S}}^{2}+\omega_{\mathsf{B}}^{2}+8\lambda^{2})/\mathsf{D}$, $K_{\chi}=8\lambda^{2}/\mathsf{D}$, with $\mathsf{D}=2(\omega_{\mathsf{S}}^{2}+\omega_{\mathsf{B}}^{2}+2\lambda^{2})^{1/2}$, which satisy Eq. \eqref{Ts-connection-main}. Note that generally $1/\widetilde{T}_{\mathsf{S}}$ and $1/\widetilde{T}_{\mathsf{B}}$ may differ for a composite system in the thermal state \cite{SM}.

\textit{Summary and Conclusions}.---Extending upon standard thermodynamics, we have obtained a closed expression for the temperature in a general nonequilibrium quantum system by taking partial derivative of the entropy with respect to the internal energy. The key to do so is to expand the density matrix in terms of a set of independent, orthonormal observables (including the system Hamiltonian) and then to identify the corresponding expectation values (including the system internal energy) as the relevant variables describing thermodynamics of the system. 

We have shown that this temperature meets several expected properties such as consistency with the Gibbs temperature for systems in equilibrium thermal state and positivity of the temperature for passive states. In addition, this temperature has been shown to have a dynamical picture. At any instant of time the state of the system can be written as an exponential which resembles a generalized Gibbs state, where temperature appears as the prefactor of the system Hamiltonian and along with other parameters in the exponent fully describe the state. When the state becomes thermal, the temperature appears to be the only parameter needed to identify the state. In the case of correlated bipartite systems, we have also stretched this framework to assign a temperature to correlation. This temperature has been defined based on variations of the mutual entropy with respect to a binding energy, which is attributed to correlations and is inaccessible to local subsystems. 

While the temperature for a single system is straightforward within our formalism, we have argued that in a composite system the temperature depends on what information is accessible. This yields two temperatures for a subsystem depending on whether we have access to local information (entropy of the local subsystem) or total information (entropy of the total system). We have shown that these temperatures are in general different due to the existence of correlation. We have obtained a universal relation which connects the temperatures of the subsystems, temperature of the composite system as a whole, and the temperature of correlation. 

\textit{Acknowledgements.}---\textit{Acknowledgements.}---We thank A. Allahverdyan and S. Marcantoni for useful discussions. This work was supported in part by the Academy of Finland's QTF Center of Excellence program (Project No. 312298), Sharif University of Technology’s Office of Vice President for Research and Technology, and the Institute for Research in Fundamental Sciences (IPM). This project/research was supported by FQXi Grant Number: FQXi-IAF19-07 from the Foundational Questions Institute Fund, donor advised fund of Silicon Valley Community Foundation.


%

\onecolumngrid
\newpage
\appendix

\section{Supplemental Material}

\section{I. Agreement of the derivative of the entropy with respect to the internal energy with the derivative of the entropy with respect to the heat}

We start from the definition of heat and work in quantum systems. A commonly used definition of heat and work has been introduced based on the change in the density matrix and the change in the Hamiltonian, respectively, as
\begin{align}
 \delta Q=\mathrm{Tr}[ d \varrho\,H];~~  \delta W=\mathrm{Tr}[\varrho\, d H].
\label{conventional-heat-work}
\end{align}
However, there are some shortcomings in these definitions of heat and work which have been discussed in Ref. \cite{entropic-based-heat-work}. In particular, according to classical thermodynamics, if entropy remains constant in a process, the corresponding energy exchange should be labeled as work \cite{book:Callen,Weimer-Mahler-WorkHeat} which is not necessarily satisfied by the definitions in Eq. \eqref{conventional-heat-work}. Accordingly, this inconvenience has led to new and thermodynamically justified definitions where heat is introduced as part of the internal energy of the open system which is associated with the entropy change and work is the rest of the energy change. Considering the spectral decomposition of the system density matrix as $\varrho=\sum_{i=1}^{d_{\mathsf{S}}} r_{i} \Pi_{i}$, where $\Pi_{i}$ is the eigenprojector of $\varrho$ related to the eigenvalue $r_{i}$, the variation of the density matrix $ d \varrho$ can be decomposed into two terms: $ \delta \varrho^{(\mathrm{ev})}=\sum_{i=1}^{d_{\mathsf{S}}}  d r_{i} \,\Pi_{i}$ which is related to the variations of the eigenvalues and $ \delta \varrho^{(\mathrm{ep})}=\sum_{i=1}^{d_{\mathsf{S}}} r_{i}  \,d \Pi_{i}$ which is related to the variations of the eigenprojectors, such that $ d \varrho= \delta \varrho^{(\mathrm{ev})}+ \delta \varrho^{(\mathrm{ep})}$. Based on this decomposition the new definitions of heat and work are given by
\begin{align}
 \delta \mathbbmss{Q}=&\mathrm{Tr}[ \delta \varrho^{(\mathrm{ev})}\,H];~~  \delta \mathbbmss{W}=\mathrm{Tr}[ \delta \varrho^{(\mathrm{ep})}H]+ \mathrm{Tr}[\varrho\, d H].
\label{entropic-heat-work}
\end{align}
In the following, starting from Eq. \eqref{classical-T}, we show that the temperatures obtained from the definitions of heat in Eqs. \eqref{conventional-heat-work} and \eqref{entropic-heat-work} indeed coincide. 
It should be noted that in cases where there is an interaction between the system and the environment, the Hamiltonian in Eqs. \eqref{conventional-heat-work} and \eqref{entropic-heat-work} is not necessarily the bare Hamiltonian $H_{0}$ of the unperturbed system, but it might be replaced by a suitable relevant effective Hamiltonian \cite{SciRep}. For example, for an open system with Markovian dynamics we may have $H=H_{0}+H_{\mathrm{Lamb} }$, where $H_{\mathrm{Lamb}}$ is an environment induced Lamb-shift correction to system Hamiltonian \cite{BreuerBook}.

\textit{Temperature by Eq. \eqref{conventional-heat-work}.---}Expanding $\varrho(t)$ and $ d \varrho(t)$ in terms of the set of observables $\{O_{i}\}_{i=0}^{d^{2}-1}$ leads to
\begin{align}
\varrho(\{ x_{i}\};t) &=\textstyle{\sum_{i=0}^{d^{2}-1}}  x_{i}(t)\, O^{(i)}(t),
\label{expansion-rho-sm} \\
d \varrho(\{ x_{i}\};t)&= \textstyle{\sum_{i=1}^{d^{2}-1}} O^{(i)}(t)\,\delta y_{i}(t),
\label{expansion-drho}
\end{align}
where $\delta y_{i}=\mathrm{Tr}[ d \varrho(\{x_{i}\};t) \,O^{(i)}]$. The exact differentials $dx_{i}$ and the infinitesimal, generically nonintegrable, variations $\delta y_{i}$ are connected through the relation $ \delta  y_{i}= d x_{i} + \sum_{j>0}  x_{j} a_{i\,\,}^{\,\,j}  d t$, where $a_{i\,\,}^{\,\,j}=\mathrm{Tr}[\partial_{t} O^{(i)} \,O^{(j)}]$. It should be noted that since $ d \varrho$ is a $d \times d$ Hermitian operator, it contains $d^{2}$ independent real parameters. Since $\mathrm{Tr}[ d \varrho\, O^{(0)}]=0$, this means that all the elements $\{ \delta  y_{i}\}_{i>0}$ contributing to the variations of the density matrix are independent. 

Now we show that the change in entropy is also given in terms of $\{ \delta y_{i}\}_{i>0}$. Starting from the definition of the von Neumann entropy $S=-\mathrm{Tr}[\varrho \log \varrho]$ and using Eq. \eqref{expansion-drho}, it is straightforward to obtain that
\begin{align}
 d S=-\mathrm{Tr}[ d \varrho \log \varrho]\equiv - \textstyle{\sum_{i>0}} \mathrm{Tr}[O^{(i)} \log \varrho]\, \delta   y_{i}. 
\label{ssaa}
\end{align}
One can also write $ \delta Q$ in terms of $ \delta  y_{i}$ by replacing Eq. \eqref{expansion-drho} into Eq. \eqref{conventional-heat-work} as
\begin{align}
 \delta Q= \textstyle{\sum_{i>0}}  \mathrm{Tr}[O^{(i)}\,H]\,\delta y_{i}\equiv h\,\delta y_{1},
\label{dQ-cov-dy1}
\end{align}
where in the last relation we have written $H$ in terms of $O^{(1)}$ from Eq. \eqref{tracelessH} and have used the orthonormality of the operators $O^{(i)}$. This relation shows that heat exchange (modified by the $1/h$ prefactor) can be considered as one of the independent infinitesimal variations $\delta y_{1}=(1/h) \delta Q$. As a result, by replacing Eq. \eqref{dQ-cov-dy1} into Eq. (\ref{ssaa}), the entropy change (denoted by $\delta S$, which is the same as $dS$) can be recast as
\begin{equation}
\delta S=-\frac{1}{h}\mathrm{Tr}[O^{(1)} \log\varrho]\, \delta Q - \textstyle{\sum_{i\geqslant 2}} \mathrm{Tr}[O^{(i)}\log\varrho]\, \delta y_{i},
\label{ds-new}
\end{equation}
from which we derive the following definition of nonequilibrium temperature:
\begin{equation}
\frac{1}{T}=\left(\frac{\delta S}{\delta Q}\right)_{ \delta y_{2}=\delta y_{3}=\ldots=0} = -\frac{1}{h}\mathrm{Tr}[O^{(1)} \log\varrho].
\label{temp-new-sup}
\end{equation}

\textit{Temperature by Eq. \eqref{entropic-heat-work}.---}Here we show that temperature will be obtained similar to Eq. \eqref{def:T} if we use the entropy-based definitions of heat and work. To do so we first note that entropy changes are essentially related to changes in the eigenvalues of the density matrix such that 
\begin{align}
 d S=-\mathrm{Tr}[ d\varrho \log \varrho]\equiv -\mathrm{Tr}[ \delta \varrho^{(\mathrm{ev})}\log \varrho].
\label{entropic-S}
\end{align}
Following similar steps as in the previous section we can decompose $ \delta \varrho^{(\mathrm{ev})}$ in terms of the chosen operator basis as
\begin{equation}
\delta \varrho^{(\mathrm{ev})}(\{ x_{i}\};t)= \textstyle{\sum_{i=1}^{d^{2}-1}} O^{(i)}(t)\,\delta z_{i}(t),
\label{expansion-drho1}
\end{equation}
where $\delta z_{i}=\mathrm{Tr}[ \delta \varrho^{(\mathrm{ev})}(t)\,O^{(i)}]$. Again, since $ \delta \varrho^{(\mathrm{ev})}$ is a $D\times D$-dimensional Hermitian operator, the infinitesimal quantities $ \delta z_{i}$ are independent variations. Replacing Eq. \eqref{expansion-drho1} in the definition of heat in Eq. \eqref{entropic-heat-work}, it can be seen that $ \delta \mathbbmss{Q}=h\, \delta z_{1}$. On the other hand, inserting Eq. \eqref{expansion-drho1} into Eq. \eqref{entropic-S} yields 
\begin{equation}
 d S=-\frac{1}{h}\mathrm{Tr}[O^{(1)} \log\varrho]\, \delta \mathbbmss{Q} - \textstyle{\sum_{i\neq 0,1}} \mathrm{Tr}[O^{(i)}\log\varrho]\, \delta z_{i},
\label{entropic-ds-new}
\end{equation}
from which temperature $T$ is obtained as in Eq. \eqref{temp-new} and coincides with the expression in Eq. \eqref{def:T}.

\section{II. Relation between $ \delta y_{i}$ and $ d  x_{i}$}

By calculating the exact differential of $\varrho(\{ x_{i}\};t)$ as $ d \varrho=\sum_{i}\partial_{x_{i}} \varrho\, d  x_{i}+\partial_t \varrho\, d t$, and using the fact that $x_{0}=1/\sqrt{d}$ and $O^{(0)}=\mathbbmss{I}/\sqrt{d}$ are time-independent, one obtains that 
\begin{equation}
\label{1}
d \varrho(\{ x_{i}\};t)= \textstyle{\sum_{i>0}}  d  x_{i} \,O^{(i)}+ \textstyle{\sum_{i>0}}  x_{i}  \,d O^{(i)} = 
\textstyle{\sum_{i>0}}  d  x_{i}\,O^{(i)} + \textstyle{\sum_{i>0}}  x_{i}\, \partial_{t} O^{(i)}\, d t.
\end{equation}
It should be noted that here $ d X$ means the exact differential of $X$, that is, $ d X=\sum_{i} \partial_{ x_{i}} X~ d  x_{i}+\partial_ t X~ d t$. Expansion of $\partial_{t} O^{(i)}$, for $i>0$, gives $\displaystyle \partial_{t} O^{(i)}= \textstyle{\sum_{j>0}} a^{i}_{j} O^{(j)}$, 
in which $a^{i}_{j}=\mathrm{Tr}[\partial_{t} O^{(i)} \,O^{(j)}]$. Inserting this into Eq. \eqref{1} gives
\begin{equation}
\label{drho(yi)}
d \varrho(\{ x_{i}\};t)= \textstyle{\sum_{i>0}} \big( d   x_{i} + \textstyle{\sum_{j>0}}  x_{j} a^{i}_{j}  d t\big) O^{(i)}=: \textstyle{\sum_{i>0}} \delta  y_{i}\, O^{(i)},
\end{equation}
whence one reads $\delta  y_{i}= \mathrm{Tr}[ d \varrho\, O^{(i)}]=d x_{i}+\textstyle{\sum_{j>0}}  x_{j} a_{i\,\,}^{\,\,j}  d t$.

\ignore{

\section{Invariance of the internal entropy production with different choices of operator basis}

One can see that the definition of $ d \Sigma$ is independent of our choice of basis operators $\{O^{(i)}\}_{i=0}^{d^{2}-1}$. Let $\{O'_{i}\}_{i=0}^{d^{2}-1}$ be another set of orthonormal operators such that still $O'_{0}=O_{0}$ and $O'_{1}=O^{(1)}$. These bases must be related through an orthogonal matrix $\mathcal{U}$ as $O^{(i)}=\sum_{k=2}^{d^{2}-1}\mathcal{U}_{ik}O'_{k}$, for $i\geqslant 2$. From Eq. (\ref{entro_prod}) one can see
\begin{align*}
 d \Sigma&= - \textstyle{\sum_{i=2}^{d^{2} -1}} \mathrm{Tr}[O^{(i)}\log \varrho]\,\mathrm{Tr}[O^{(i)} d \varrho]\\
&=- \textstyle{\sum_{i=2}^{d^{2} -1}} \textstyle{\sum_{k=2}^{d^{2} -1}} \textstyle{\sum_{l=2}^{d^{2} -1}} \mathcal{U}_{ik}\mathcal{U}_{il}\mathrm{Tr}[O'_{k}\log \varrho]\,\mathrm{Tr}[O'_{l} d \varrho]\\
&= -  \textstyle{\sum_{k=2}^{d^{2} -1}} \textstyle{\sum_{l=2}^{d^{2} -1}} \big( \textstyle{\sum_{i=2}^{d^{2} -1}} (\mathcal{U}^{T})_{ki}(\mathcal{U})_{il} \big)\mathrm{Tr}[O'_{k}\log \varrho]\,\mathrm{Tr}[O'_{l} d \varrho]\\
&=- \textstyle{\sum_{k=2}^{d^{2}-1}} \mathrm{Tr}[O'_{k}\log \varrho]\,\mathrm{Tr}[O'_{k} d \varrho].
\end{align*}

}
\section{III. Invariance of the temperature with different choices of operator basis}

Here we show that the relation 
\begin{equation}
S+\frac{1}{h}\mathrm{Tr}[O^{(1)} \log\varrho]\, U + I =\textstyle{\sum_{i\geqslant 2}} \mathrm{Tr}[O^{(i)}\log\varrho]\, x_{i},
\end{equation}
where $\mathcal{I}$ denotes irrelevant terms, is independent of our choice of basis operators $\{O^{(i)}\}_{i=0}^{d^{2}-1}$. As long as $O^{(0)}$ and $O^{(1)}$ are in the set of the chosen operators we obtain the same value for the partial derivative of $S$ with respect to $U$, that is, we obtain the same value for the temperature. To see this, let $\{O'_{i}\}_{i=0}^{d^{2}-1}$ be another set of orthonormal operators such that still $O'_{0}=O^{(0)}$ and $O'_{1}=O^{(1)}$. These bases need to be related through a unitary matrix $\mathcal{U}$ as $O^{(i)}=\sum_{k=2}^{d^{2}-1}\mathcal{U}_{ik}O'_{k}$, for $i\geqslant 2$. Thus,
\begin{align*}
\textstyle{\sum_{i\geqslant 2}} \mathrm{Tr}[O^{(i)}\log\varrho]\, x_{i}&= - \textstyle{\sum_{i\geqslant 2}}\mathrm{Tr}[O^{(i)}\log \varrho]\,\mathrm{Tr}[O^{(i)} \varrho]\\
&=- \textstyle{\sum_{i\geqslant 2}} \textstyle{\sum_{k\geqslant 2}} \textstyle{\sum_{l\geqslant 2}} \mathcal{U}_{ik}\mathcal{U}_{il}\mathrm{Tr}[O'_{k}\log \varrho]\,\mathrm{Tr}[O'_{l} \varrho]\\
&= -  \textstyle{\sum_{k\geqslant 2}} \textstyle{\sum_{l\geqslant 2}} \left( \textstyle{\sum_{i\geqslant 2}} (\mathcal{U}^{T})_{ki}(\mathcal{U})_{il} \right)\mathrm{Tr}[O'_{k}\log \varrho]\,\mathrm{Tr}[O'_{l} \varrho]\\
&=- \textstyle{\sum_{k\geqslant 2}} \mathrm{Tr}[O'_{k}\log \varrho]\,\mathrm{Tr}[O'_{k} \varrho]\nonumber\\
&=- \textstyle{\sum_{k\geqslant 2}} \mathrm{Tr}[O'_{k}\log \varrho]\,x'_k.
\end{align*}
It is interesting to note that $S + (1/h)\mathrm{Tr}[O^{(1)} \log\varrho]\, U \equiv S-U/T=-F/T$, in which $F$ is the Helmholtz free energy. Thus, the Helmholtz free energy in terms of the variables of the system is given by 
\begin{align}
F&=T\textstyle{\sum_{i\geqslant 2}} \mathrm{Tr}[O^{(i)}\log\varrho]\, x_{i}+ I\nonumber\\
&=\frac{\Delta^{2} H}{\mathrm{Cov}(H,\mathbbmss{H})}\textstyle{\sum_{i\geqslant 2}} \mathrm{Tr}[O^{(i)}\log\varrho]\, \mathrm{Tr}[O^{(i)} \varrho]+\mathcal{I}
\end{align}

\section{IV. A useful relation}

\begin{lemma}
\label{lemma:1}
Let $A$ be a traceless Hermitian matrix and $B$ be a positive-definite matrix. Assume that $A$ has the spectral decomposition $A=\sum_{i\in I\equiv \{1,\ldots,n\}} a_{i} |a_{i}\rangle\langle a_{i}|$, where the eigenvalues are ordered increasingly ($a_{i}\geqslant a_{i-1},\,\forall i$), and $b_{i}=\langle a_{i}|B|a_{i}\rangle$s also have the same ordering ($b_{i}\geqslant b_{i-1},\,\forall i$). Then 
\begin{equation}
\mathrm{Tr}[AB]\geqslant 0.
\end{equation}
\end{lemma}

\textit{Proof}. Let $a^{(+)}_{i}$, $i\in I^{(+)}$ denote positive eigenvalues of $A$ and $-a^{(-)}_{i}$, $i\in I^{(-)}$, with $a^{(-)}_{i}\geqslant 0$, negative ones. Note that $\mathrm{Tr}[A]=\sum_{i\in I^{(+)}} a^{(+)}_{i} - \sum_{i \in I^{(-)}} a^{(-)}_{i}=0$. In addition, let us denote by $b^{(+)}_{i}$, for $i\in I^{(+)}$, and by $b^{(-)}_{i}$, for $i\in I^{(-)}$, the positive mean-values of $B$ with respect to the eigenvectors of $A$ corresponding to positive and negative eigenvalues, respectively. Thus, due to the ordering property of $B$ we have $b^{(+)}_{i}\geqslant b^{(-)}_{j}$, $\forall i,j$. Now we can write
\begin{align*}
\mathrm{Tr}[AB] &= \textstyle{\sum_{i}} a_{i} b_{i} = \textstyle{\sum_{i\in I^{(+)}}} a^{(+)}_{i} b^{(+)}_{i} - \textstyle{\sum_{i\in I^{(-)}}} a^{(-)}_{i} b^{(-)}_{i} \geqslant \min_{i\in I^{(+)}} b^{(+)}_{i} \textstyle{\sum_{i\in I^{(+)}}} a^{(+)}_{i}  - \max_{i\in I^{(-)}} b^{(-)}_{i}  \textstyle{\sum_{i \in I^{(-)}}} a^{(-)}_{i} \\
&= \Big( \min_{i\in I^{(+)}} b^{(+)}_{i} -\max_{i\in I^{(-)}} b^{(-)}_{i} \Big) \textstyle{\sum_{i\in I^{(+)}}} a^{(+)}_{i} \geqslant 0. \hskip10.5cm \square
\end{align*}

An alternative proof of the positivity of the instantaneous temperature for passive states goes as follows. Consider a quantum system with effective Hamiltonian $H_{\mathsf{S}}(t)$ whose state $\varrho_{\mathsf{S}}(t)$ satisfies the passivity condition, i.e.,
\begin{align}
\label{passive}
H_{\mathsf{S}}(t)= \textstyle{\sum_{i}} E_{i} \ketbra{\epsilon_{i}}{\epsilon_{i}} ~~~, ~~ E_{i+1}\geqslant  E_{i}\\
\varrho_{\mathsf{S}}(t)= \textstyle{\sum_{i}} \eta_{i} \ketbra{\epsilon_{i}}{\epsilon_{i}} ~~~, ~~ \eta_{i+1}\leqslant \eta_{i},
\end{align}
{where, for simplicity and ease of notation, on the right-hand sides we have omitted the time-dependence of eigenvalues and eigenvectors. The numerator of Eq. \eqref{def:T} is positive; indeed, we have
\begin{align}
d \times \,\mathrm{Tr}[H_{\mathsf{S}}(t) \log{\varrho_{\mathsf{S}}(t)}]-\mathrm{Tr}[H_{\mathsf{S}}(t)]\, \mathrm{Tr}[\log{\varrho_{\mathsf{S}}(t)}]&= d \times \textstyle{\sum_{i}} E_{i} \log \eta_{i} - \textstyle{\sum_{i,j}} E_{i}  \log \eta_{j}\nonumber\\
&=\frac{1}{2} \textstyle{\sum_{ij}} \left(E_{i} \log \eta_{i}+E_{j} \log \eta_{j} - E_{i} \log \eta_{j} - E_{j} \log \eta_{i}\right)\nonumber \\
\label{pass-ineq}
&=\frac{1}{2} \textstyle{\sum_{ij}} (E_{i}-E_{j}) (\log \eta_{i} - \log \eta_{j}),
\end{align}
where $\sum_{i,j=1}^{d} E_{i} \log \eta_{i}= d \times \sum_{i=1}^{d} E_{i} \log \eta_{i}$. The expression in Eq. \eqref{pass-ineq} is positive since the passivity condition (\ref{passive}) yields
\begin{align}
i>j&: ~ E_{i}>E_{j}\Rightarrow ~ E_{i}-E_{j}>0 \nonumber\\
i>j&:~\eta_{i}<\eta_{j}~ \mathrm{and} ~0\leqslant \eta_{i},\eta_{j} \leqslant 1  \Rightarrow \frac{1}{\eta_{i}}\geqslant  \frac{1}{\eta_{j}}\Rightarrow -\log \eta_{i}\geqslant  -{\log \eta_{j}}.
\end{align}
The denominator of Eq. \eqref{def:T} and thus the temperature is also positive; indeed,
\begin{align}
\Delta^{2}H_{\mathsf{S}}^{(\mathrm{eff})}=d\times\,\mathrm{Tr}[(H_{\mathsf{S}}^{(\mathrm{eff})})^{2}]-\mathrm{Tr}[H_{\mathsf{S}}^{(\mathrm{eff})}]^{2}=d\times \textstyle{\sum_{i}} E_{i}^{2} - \textstyle{\sum_{i,j}} E_{i} E_{j}= \textstyle{\sum_{ij}} (E_{i}^{2}+E_{j}^{2}-2E_{i} E_{j})= \textstyle{\sum_{ij}} (E_{i}-E_{j})^{2}\geqslant  0.
\end{align}

In the main text we have used this lemma to prove the positivity of the temperature for passive states.

\section{V. Details of the example}

Setting $\Delta_{\pm}=(\omega_{\mathsf{S}} \pm \omega_{\mathsf{B}})/2$, the eigenvalues of $H_{\mathsf{SB}}$ are 
\begin{equation}
\label{2qubits2a}
E_{1}=\Delta_{+}\ , \ E_{2}=\eta\ ,\ E_{3}=-\eta\ ,\ E_4=-\Delta_{+},
\end{equation}
with $\eta=\sqrt{\Delta_{-}^{2}+16\lambda^{2}}$. The corresponding eigenvectors read
\begin{gather}
\label{2qubits2a}
\vert E_{1}\rangle=\vert 00\rangle\ ,\ \vert E_{2}\rangle=\zeta_{+}\vert 01\rangle+\zeta_{-}\vert 10\rangle\\
\label{2qubits2b} 
\vert E_{3}\rangle=\zeta_{-}\vert 01\rangle-\zeta_{+}\vert 10\rangle\ ,\ \vert E_4\rangle=\vert 11\rangle\ ,
\end{gather}
where $\zeta{_\pm}=\sqrt{(1/2) [1\pm \Delta_{-}/\eta]}$. 

\section{VI. Connection between the temperature of the total system and the temperatures of the subsystems}

We define $O_{\chi}$ as below such that it is orthonormal to $O_{\mathsf{S}}\otimes \mathbbmss{I}_{\mathsf{B}}$ and $\mathbbmss{I}_{\mathsf{S}} \otimes O_{\mathsf{B}}$,
\begin{align}
O_{\chi} =\frac{1}{h_{\chi}}\left(O_{\mathrm{I}}-\frac{\mathrm{Tr}[O_{\mathrm{I}} \, O_{\mathsf{S}} ]}{d_{\mathsf{B}}}\, O_{\mathsf{S}}  \otimes \mathbbmss{I}_{\mathsf{B}}-\frac{\mathrm{Tr}[O_{\mathrm{I}} \, O_{\mathsf{B}}]}{d_{\mathsf{S}}}\, \mathbbmss{I}_{\mathsf{S}} \otimes O_{\mathsf{B}} \right),
\label{O-chi}
\end{align}
where $h_{\chi}=\sqrt{1-\frac{\mathrm{Tr}[O_{\mathrm{I}} \, O_{\mathsf{S}}]^{2}}{d_{\mathsf{B}}}-\frac{\mathrm{Tr}[O_{\mathrm{I}} \, O_{\mathsf{B}}]^{2}}{d_{\mathsf{S}} }}$ is the normalization factor (and we have used $\mathrm{Tr}[O_{\mathrm{I}}^{2}]=1$).
 The normalized traceless operator obtained from the Hamiltonian of the total system is given by $O^{(1)}_{\mathsf{SB}}=\frac{1}{h_{\mathsf{SB}}}(H_{\mathsf{SB}}-\mathrm{Tr}[H_{\mathsf{SB}}]\frac{\mathbbmss{I}_{\mathsf{SB}}}{d_{\mathsf{SB}} })$. By expanding $O^{(1)}_{\mathsf{SB}}$ in terms of $O_{\mathsf{S}}$, $O_{\mathsf{B}}$, and $O_{\mathsf{\chi}}$ we have
\begin{align}
O^{(1)}_{\mathsf{SB}}=C_{\mathsf{S}}  \, O_{\mathsf{S}}\otimes \mathbbmss{I}_{\mathsf{B}} + C_{\mathsf{B}} \, \mathbbmss{I}_{\mathsf{S}} \otimes O_{\mathsf{B}} + C_{\chi} \, O_{\chi},
\label{O-sb-expand}
\end{align}
where 
\begin{align}
C_{\mathsf{S}} &=\mathrm{Tr}[O^{(1)}_{\mathsf{SB}} \, O_{\mathsf{S}}]/d_{\mathsf{B}}=\frac{1}{h_{\mathsf{SB}}  h_{\mathsf{S}}  d_{\mathsf{B}}}(\mathrm{Tr}[H_{\mathsf{SB}} \,  H_{\mathsf{S}}^{(\mathrm{eff})} ]-\mathrm{Tr}[H_{\mathsf{SB}} ]\, \mathrm{Tr}[ H_{\mathsf{S}}^{(\mathrm{eff})} ]/d_{\mathsf{S}})
\label{C_S}
\\
C_{\mathsf{B}} &=\mathrm{Tr}[O^{(1)}_{\mathsf{SB}}\, O_{\mathsf{B}}]/d_{\mathsf{S}}=\frac{1}{h_{\mathsf{SB}}  h_{\mathsf{B}}  d_{\mathsf{S}}}(\mathrm{Tr}[H_{\mathsf{SB}}  \,  H_{\mathsf{B}}^{(\mathrm{eff})} ]-\mathrm{Tr}[H_{\mathsf{SB}} ]\, \mathrm{Tr}[ H_{\mathsf{B}}^{(\mathrm{eff})} ]/d_{\mathsf{B}})\\
C_{\chi}&=\mathrm{Tr}[O^{(1)}_{\mathsf{SB}} \, O_{\chi}]\nonumber\\
&=\frac{1}{h_{\chi}}\mathrm{Tr}[O^{(1)}_{\mathsf{SB}}  \, O_\mathrm{I}]-\frac{1}{h_{\chi} d_{\mathsf{B}}} \mathrm{Tr}[O^{(1)}_{\mathsf{SB}}  \, O^{(1)}_{\mathsf{S}} ] \, \mathrm{Tr}[O_{\mathrm{I}}  \,O^{(1)}_{\mathsf{S}} ]
- \frac{1}{h_{\chi}  d_{\mathsf{S}}}\mathrm{Tr}[O^{(1)}_{\mathsf{SB}}  \,O^{(1)}_{\mathsf{B}} ]\, \mathrm{Tr}[O_{\mathrm{I}}  \,O^{(1)}_{\mathsf{B}} ].
\label{C_chi}
\end{align}
The terms in $C_{\chi}$ are given by
\begin{align}
\mathrm{Tr}[O_{\mathrm{I}}  \, O_{\mathsf{S}} ]&=\frac{1}{h_{\mathrm{I}}  h_{\mathsf{S}} }(\mathrm{Tr}[H_{\mathrm{I}}  \,  H_{\mathsf{S}}^{(\mathrm{eff})} ]-\mathrm{Tr}[H_\mathrm{I}]\, \mathrm{Tr}[ H_{\mathsf{S}}^{(\mathrm{eff})} ]/d_{\mathsf{S}}),\\
\mathrm{Tr}[O_{\mathrm{I}}  \, O_{\mathsf{B}} ]&=\frac{1}{h_{\mathrm{I}}  h_{\mathsf{B}} }(\mathrm{Tr}[H_{\mathrm{I}}  \,  H_{\mathsf{B}}^{(\mathrm{eff})} ]-\mathrm{Tr}[H_\mathrm{I}]\, \mathrm{Tr}[ H_{\mathsf{B}}^{(\mathrm{eff})} ]/d_{\mathsf{B}}).
\end{align}

Two other operators $O^{(2)}_{\mathsf{SB}}$ and $O^{(3)}_{\mathsf{SB}}$ together with $O^{(1)}_{\mathsf{SB}}$ will form another complete set of operator basis for the subspace $\mathpzc{L}_{H}=\mathrm{span}\{O_{\chi}, O_{\mathsf{S}}\otimes \mathbbmss{I}_{\mathsf{B}}, \mathbbmss{I}_{\mathsf{S}} \otimes O_{\mathsf{B}}\}\equiv \mathrm{span}\{O_{\mathrm{I}}, O_{\mathsf{S}}\otimes \mathbbmss{I}_{\mathsf{B}}, \mathbbmss{I}_{\mathsf{S}} \otimes O_{\mathsf{B}}\}$ whereto the total Hamiltonian belongs.  The operators $O^{(2)}_{\mathsf{SB}}$ and $O^{(3)}_{\mathsf{SB}}$ can be obtained as
\begin{gather}
O^{(2)}_{\mathsf{SB}}=\frac{C_{\mathsf{B}} }{d_{\mathsf{B}} } O_{\mathsf{S}}\otimes \mathbbmss{I}_{\mathsf{B}}  -\frac{C_{\mathsf{S}} }{d_{\mathsf{S}} } \mathbbmss{I}_{\mathsf{S}}  \otimes O_{\mathsf{B}},
\label{eq1}
\\
O^{(3)}_{\mathsf{SB}}=C_{\mathsf{S}}  O_{\mathsf{S}}\otimes \mathbbmss{I}_{\mathsf{B}} + C_{\mathsf{B}}\mathbbmss{I}_{\mathsf{S}}  \otimes O_{\mathsf{B}}-\frac{C_{\mathsf{S}}^{2} d_{\mathsf{B}} +C_{\mathsf{B}}^{2} d_{\mathsf{S}} }{C_{\chi}} O_{\chi}.
\label{eq2}
\end{gather}
Having this set of observables in addition to the set of remainder observables $\{R^{(i)}\}_{i=1}^{d_{\mathsf{SB}}-4}$ chosen to be orthogonal to the subspace $\mathpzc{L}_{H}$, we have a complete set of observables $\{\mathbbmss{I}_{\mathsf{SB}}, O^{(i)}_{\mathsf{SB}}, R^{(i)}\}_{i}$ forming a basis for the total space of observables. By expanding $\varrho_{\mathsf{SB}} $ we get to
\begin{align}
\varrho_{\mathsf{SB}} = \textstyle{\sum_{i=1}^{3}} u_{i}\, O^{(i)}_{\mathsf{SB}}+\sum_{i=1}^{d_{\mathsf{SB}}-4} r_{i} R^{(i)},
\end{align}
in which we have defined the related set of independent variables in expanding the state of the total system as $u_{i}=\mathrm{Tr}[\varrho_{\mathsf{SB}}  \, O^{(i)}_{\mathsf{SB}}]$ and $r_{i}=\mathrm{Tr}[\varrho_{\mathsf{SB}} R^{(i)}]$, where $u_{1}=(U_{\mathsf{SB}} -\mathrm{Tr}[H_{\mathsf{SB}}])/h_{\mathsf{SB}} $. From the relation for the entropy in the total system we have
\begin{align}
S_{\mathsf{SB}}&=-\mathrm{Tr}[\varrho_{\mathsf{SB}}\log \varrho_{\mathsf{SB}}]\nonumber\\ 
&=-\frac{U_{\mathsf{SB}}-\mathrm{Tr}[H_{\mathsf{SB}}]}{h_{\mathsf{SB}}}\,\mathrm{Tr}[O^{(1)}_{\mathsf{SB}}\,\log \varrho_{\mathsf{SB}}]+ \textstyle{\sum_{i=2}^{3}} u_{i} \,\mathrm{Tr}[O^{(i)}_{\mathsf{SB}}\,\log \varrho_{\mathsf{SB}}]+ \textstyle{\sum_{i=1}^{d_{\mathsf{SB}}-4}} r_{i}\,\mathrm{Tr}[R^{(i)}\,\log \varrho_{\mathsf{SB}}],
\label{tot-entrop-expan}
\end{align}
hence 
\begin{align}
\frac{1}{T_{\mathsf{SB}}}=\left(\frac{\partial S_{\mathsf{SB}}}{\partial U_{\mathsf{SB}}}\right)_{u_{2},u_{3},\{r_{i}\}}=-\frac{1}{h_{\mathsf{SB}}}\,\mathrm{Tr}[O^{(1)}_{\mathsf{SB}}\,\log \varrho_{\mathsf{SB}}].
\end{align}
To obtain the connection between the temperature of the subsystems, the correlation temperature, and the temperature of the total system, we start from 
\begin{align}
S_{\mathsf{SB}}=S_{\mathsf{S}}+S_{\mathsf{B}}-S_{\chi},
\label{StotvsSsSbSchi}
\end{align}
and take partial derivative of both sides of the above relation with respect to (i) $U_{\chi}$ and (ii) $U_{\mathsf{SB}}$.

 \textit{Derivative with respect to $U_{\chi}$.}---Taking derivative of Eq. \eqref{StotvsSsSbSchi} with respect to $U_{\chi}$ gives
\begin{gather}
\left(\frac{\partial S_{\mathsf{SB}}}{\partial U_{\chi}}\right)_{U_{\mathsf{S}} ,U_{\mathsf{B}} ,\{r_{i}\}}
=
\left(\frac{\partial S_{\mathsf{S}}}{\partial U_{\chi}}\right)_{U_{\mathsf{S}} ,U_{\mathsf{B}} ,\{r_{i}\}}
+
\left(\frac{\partial S_{\mathsf{B}}}{\partial U_{\chi}}\right)_{U_{\mathsf{S}} ,U_{\mathsf{B}} ,\{r_{i}\}}
-
\left(\frac{\partial S_{\chi}}{\partial U_{\chi}}\right)_{U_{\mathsf{S}} ,U_{\mathsf{B}} ,\{r_{i}\}},\nonumber
\\
\left(\frac{\partial S_{\mathsf{SB}}}{\partial u_{1}}\right)_{u_{2},u_{3},\{r_{i}\}}
\left(\frac{\partial u_{1}}{\partial U_{\chi}}\right)_{U_{\mathsf{S}}, U_{\mathsf{B}} ,\{r_{i}\}}
+
\left(\frac{\partial S_{\mathsf{SB}}}{\partial u_{2}}\right)_{u_{1},u_{3},\{r_{i}\}}
\left(\frac{\partial u_{2}}{\partial U_{\chi}}\right)_{U_{\mathsf{S}} ,U_{\mathsf{B}} ,\{r_{i}\}}
+
\left(\frac{\partial S_{\mathsf{SB}}}{\partial u_{3}}\right)_{u_{1},u_{2},\{r_{i}\}}
\left(\frac{\partial u_{3}}{\partial U_{\chi}}\right)_{U_{\mathsf{S}} ,U_{\mathsf{B}} ,\{r_{i}\}}
=0+0-\frac{1}{T_{\chi}},\nonumber
\\
\frac{h_{\mathsf{SB}}}{T_{\mathsf{SB}}} \times \frac{1}{h_{\mathsf{SB}}}
+0
+\left(\frac{\partial S_{\mathsf{SB}}}{\partial u_{3}}\right)_{u_{1},u_{2},\{r_{i}\}}
\left(\frac{\partial u_{3}}{\partial U_{\chi}}\right)_{U_{\mathsf{S}} ,U_{\mathsf{B}} ,\{r_{i}\}}
=-\frac{1}{T_{\chi}},\nonumber
\\
\frac{1}{T_{\mathsf{SB}}}+\left(\frac{\partial S_{\mathsf{SB}}}{\partial u_{3}}\right)_{u_{1},u_{2},\{r_{i}\}}
\left(\frac{\partial u_{3}}{\partial U_{\chi}}\right)_{U_{\mathsf{S}} ,U_{\mathsf{B}} ,\{r_{i}\}}=-\frac{1}{T_{\chi}}
\end{gather}

\textit{Derivative with respect to $U_{\mathsf{SB}}$.---}Taking derivative with respect to $U_{\mathsf{SB}}$ gives 
 \begin{align}
\left(\frac{\partial S_{\mathsf{SB}}}{\partial U_{\mathsf{SB}}}\right)_{u_{2},u_{3},\{r_{i}\}}
=&
\left(\frac{\partial S_{\mathsf{S}}}{\partial U_{\mathsf{S}}}\right)_{U_{\mathsf{B}},U_{\chi}}
\left(\frac{\partial U_{\mathsf{S}}}{\partial U_{\mathsf{SB}}}\right)_{u_{2},u_{3},\{r_{i}\}}
+
\left(\frac{\partial S_{\mathsf{B}}}{\partial U_{\mathsf{B}}}\right)_{U_{\mathsf{S}},U_{\chi}}
\left(\frac{\partial U_{\mathsf{B}}}{\partial U_{\mathsf{SB}}}\right)_{u_{2},u_{3},\{r_{i}\}} -
\left(\frac{\partial S_{\chi}}{\partial U_{\mathsf{S}}}\right)_{U_{\mathsf{B}},U_{\chi}}
\left(\frac{\partial U_{\mathsf{S}}}{\partial U_{\mathsf{SB}}}\right)_{u_{2},u_{3},\{r_{i}\}}
\nonumber\\ &-
\left(\frac{\partial S_{\chi}}{\partial U_{\mathsf{B}}}\right)_{U_{\mathsf{S}},U_{\chi}}
\left(\frac{\partial U_{\mathsf{B}}}{\partial U_{\mathsf{SB}}}\right)_{u_{2},u_{3},\{r_{i}\}} -\left(\frac{\partial S_{\chi}}{\partial U_{\chi}}\right)_{U_{\mathsf{S}},U_{\mathsf{B}}}
\left(\frac{\partial U_{\chi}}{\partial U_{\mathsf{SB}}}\right)_{u_{2},u_{3},\{r_{i}\}}
\label{eq75}
\end{align}
where we have used the relations $\left(\frac{\partial S_{\mathsf{S}}}{\partial U_{\mathsf{B}}}\right)_{U_{\mathsf{S}},U_{\chi}}=\left(\frac{\partial S_{\mathsf{S}}}{\partial U_{\chi}}\right)_{U_{\mathsf{S}},U_{\mathsf{B}}}=0$ and similarly $\left(\frac{\partial S_{\mathsf{B}}}{\partial U_{\mathsf{S}}}\right)_{U_{\mathsf{B}},U_{\chi}}=\left(\frac{\partial S_{\mathsf{B}}}{\partial U_{\chi}}\right)_{U_{\mathsf{S}},U_{\mathsf{B}}}=0$. Multiplying both sides of the above equation by $\left(\frac{\partial U_{\mathsf{SB}}}{\partial U_{\mathsf{S}}}\right)_{u_{2},u_{3},\{r_{i}\}}$ we can rewrite it equivalently as 
\begin{align}
\left(\frac{\partial S_{\mathsf{SB}}}{\partial U_{\mathsf{SB}}}\right)_{u_{2},u_{3},\{r_{i}\}}
\left(\frac{\partial U_{\mathsf{SB}}}{\partial U_{\mathsf{S}}}\right)_{u_{2},u_{3},\{r_{i}\}}
=
\left(\frac{\partial (S_{\mathsf{S}}-S_{\chi})}{\partial U_{\mathsf{S}}}\right)_{U_{\mathsf{B}},U_{\chi}}
+
\left(\frac{\partial (S_{\mathsf{B}}-S_{\chi})}{\partial U_{\mathsf{B}}}\right)_{U_{\mathsf{S}},U_{\chi}}
\left(\frac{\partial U_{\mathsf{B}}}{\partial U_{\mathsf{S}}}\right)_{u_{2},u_{3},\{r_{i}\}}
-
\left(\frac{\partial S_{\chi}}{\partial U_{\chi}}\right)_{U_{\mathsf{S}},U_{\mathsf{B}}}
\left(\frac{\partial U_{\chi}}{\partial U_{\mathsf{SB}}}\right)_{u_{2},u_{3},\{r_{i}\}},
\label{Ts-connections}
\end{align}
which can be written equivalently as 
\begin{align}
\frac{1}{T_{\mathsf{SB}}}
\left(\frac{\partial U_{\mathsf{SB}}}{\partial U_{\mathsf{S}}}\right)_{u_{2},u_{3},\{r_{i}\}}
=
\frac{1}{\widetilde{T}_{\mathsf{S}}}
+
\frac{1}{\widetilde{T}_{\mathsf{B}}}
\left(\frac{\partial U_{\mathsf{B}}}{\partial U_{\mathsf{S}}}\right)_{u_{2},u_{3},\{r_{i}\}}
-
\frac{1}{T_{\chi}}
\left(\frac{\partial U_{\chi}}{\partial U_{\mathsf{S}}}\right)_{u_{2},u_{3},\{r_{i}\}},
\label{Ts-connection-2}
\end{align}
where 
\begin{align}
&\left(\frac{\partial (S_{\mathsf{S}}-S_{\chi})}{\partial U_{\mathsf{S}}}\right)_{U_{\mathsf{B}},U_{\chi}} :=
 \frac{1}{\widetilde{T}_{\mathsf{S}}}=
\frac{1}{T_{\mathsf{S}}}-\left(\frac{\partial S_{\chi}}{\partial U_{\mathsf{S}}}\right)_{U_{\mathsf{B}},U_{\chi}},
\\
&\left(\frac{\partial (S_{\mathsf{B}}-S_{\chi})}{\partial U_{\mathsf{B}}}\right)_{U_{\mathsf{S}},U_{\chi}}:=
\frac{1}{\widetilde{T}_{\mathsf{B}}}=
\frac{1}{T_{\mathsf{B}}}-\left(\frac{\partial S_{\chi}}{\partial U_{\mathsf{B}}}\right)_{U_{\mathsf{S}},U_{\chi}}.
\label{Tglobal-Tlocal}
\end{align}
The last terms on the right-hand sides of the above equations can be calculated explicitly from Eq. (\ref{Schi-expanded}) of the main text. It should be noted that $S_{\mathsf{S}}-S_{\chi}\equiv S_{\mathsf{SB}}-S_{\mathsf{B}}$ and $S_{\mathsf{B}}-S_{\chi} \equiv S_{\mathsf{SB}}-S_{\mathsf{S}}$. Hence it is concluded that 
\begin{align}
\left(\frac{\partial S_{\mathsf{SB}}}{\partial U_{\mathsf{S}}}\right)_{U_{\mathsf{B}},U_{\chi}} :=&
 \frac{1}{\widetilde{T}_{\mathsf{S}}},
 \\
 \left(\frac{\partial S_{\mathsf{SB}}}{\partial U_{\mathsf{B}}}\right)_{U_{\mathsf{S}},U_{\chi}} :=&
 \frac{1}{\widetilde{T}_{\mathsf{B}}}.
\end{align}
From these relations we can conclude that $\widetilde{T}_{\mathsf{S}}$ and $\widetilde{T}_{\mathsf{B}}$ are the temperatures assigned to the subsystems by the observer of the global (correlated) system, whereas $T_{\mathsf{S}}$ and $T_{\mathsf{B}}$ are the temperatures assigned to the subsystems by the local and independent (uncorrelated) observers. 

It is also worth noting that 
\begin{align}
\left(\frac{\partial S_{\chi}}{\partial U_{\chi}}\right)_{U_{\mathsf{S}},U_{\mathsf{B}}}=\left(\frac{\partial S_{\mathsf{SB}}}{\partial U_{\chi}}\right)_{U_{\mathsf{S}},U_{\mathsf{B}}},
\end{align}
which means $\widetilde{T}_{\chi}=T_{\chi}$. Evidently we also have $\widetilde{T}_{\mathsf{SB}}=T_{\mathsf{SB}}$, hence all temperatures in Eq. \eqref{Ts-connection-2} are obtained from the point of view of the observer of the global system. 

To obtain the coefficients of the temperatures in Eq. \eqref{Ts-connection-2}, since all are partial derivatives when $u_{2}$, $u_{3}$, and $\{r_{i}\}$ are constant, we should first obtain the conditions imposed on the energies by the assumptions that these variables are intact. From Eqs. \eqref{eq1} and \eqref{eq2}, it is seen that keeping $u_{2}$ and $u_{3}$ constant means, respectively, that
\begin{gather}
 C_{\mathsf{B}} \,\mathrm{Tr}[\varrho_{\mathsf{S}}  \,O_{\mathsf{S}}]-C_{\mathsf{S}}  \mathrm{Tr}[\varrho_{\mathsf{B}}  \,O_{\mathsf{B}}]=u_{2},
\label{cond1}\\
C_{\mathsf{S}}  \mathrm{Tr}[\varrho_{\mathsf{S}}  \, O_{\mathsf{S}}]+C_{\mathsf{B}}  \mathrm{Tr}[\varrho_{\mathsf{B}}  \, O_{\mathsf{B}}]-\frac{C_{\mathsf{S}}^{2} d_{\mathsf{B}} +C_{\mathsf{B}}^{2} d_{\mathsf{S}} }{C_{\chi}} \mathrm{Tr}[\varrho_{\mathsf{SB}}  \, O_{\chi}]=u_{3}.
\label{cond2-1}
\end{gather}
Replacing Eq. \eqref{cond1} into Eq. \eqref{cond2-1} gives 
\begin{align}
\mathrm{Tr}[\varrho_{\mathsf{SB}}  \,O_{\chi}]&-\frac{C_\chi}{C_{\mathsf{S}} } \left(\frac{C_{\mathsf{S}}^{2}+C_{\mathsf{B}}^{2}}{C_{\mathsf{S}}^{2} d_{\mathsf{B}} + C_{\mathsf{B}}^{2} d_{\mathsf{S}} }\right)
\mathrm{Tr}[\varrho_{\mathsf{S}}  \, O_{\mathsf{S}}]=k_{1},
\label{cond2-2}
\end{align}
where $k_{1}$ is a constant. Using Eq. \eqref{O-chi} into the above equation we obtain
\begin{align}
\mathrm{Tr}[\varrho_{\mathsf{SB}}  \, O_{\mathrm{I}}]&-\left[\frac{C_\chi h_{\chi} }{C_{\mathsf{S}} }\left(\frac{C_{\mathsf{S}}^{2}+C_{\mathsf{B}}^{2}}{C_{\mathsf{S}}^{2} d_{\mathsf{B}} +C_{\mathsf{B}}^{2} d_{\mathsf{S}} }\right)+\frac{1}{d_{\mathsf{B}}}\mathrm{Tr}[O_{\mathrm{I}} \,O_{\mathsf{S}} ]+\frac{C_{\mathsf{B}} }{C_{\mathsf{S}} }\frac{1}{d_{\mathsf{S}}}\mathrm{Tr}[O_{\mathrm{I}} \,O_{\mathsf{B}} ]\right] \mathrm{Tr}[\varrho_{\mathsf{S}}  \,O_{\mathsf{S}}]
=k_{2},
\label{cond2}
\end{align}
where $k_{2}$ is constant. Using Eqs. \eqref{cond1} -- \eqref{cond2} yields
\begin{align}
\left(\frac{\partial U_{\mathsf{SB}}}{\partial U_{\mathsf{S}}}\right)_{u_{2},u_{3},\{r_{i}\}}&=\frac{h_{\mathsf{SB}}}{C_{\mathsf{S}} h_{\mathsf{S}}}\left[C_{\mathsf{S}}^{2}+C_{\mathsf{B}}^{2}+ C_\chi^{2} \left(\frac{C_{\mathsf{S}}^{2} + C_{\mathsf{B}}^{2}}{C_{\mathsf{S}}^{2} d_{\mathsf{B}} +C_{\mathsf{B}}^{2} d_{\mathsf{S}} }\right) \right],
\label{USB/US}\\
\left(\frac{\partial U_{\mathsf{B}}}{\partial U_{\mathsf{S}}}\right)_{u_{2},u_{3},\{r_{i}\}}&=\frac{C_{\mathsf{B}} h_{\mathsf{B}}}{C_{\mathsf{S}} h_{\mathsf{S}}},
\label{UB/US}\\
\left(\frac{\partial U_{\chi}}{\partial U_{\mathsf{S}}}\right)_{u_{2},u_{3},\{r_{i}\}}&=\frac{h_{\mathrm{I}}}{h_{\mathsf{S}} C_{\mathsf{S}}}\left[C_\chi h_{\chi} \left(\frac{C_{\mathsf{S}}^{2} + C_{\mathsf{B}}^{2}}{C_{\mathsf{S}}^{2} d_{\mathsf{B}} +C_{\mathsf{B}}^{2} d_{\mathsf{S}} }\right)+\frac{C_{\mathsf{S}}}{d_{\mathsf{B}}}\mathrm{Tr}[O_{\mathrm{I}} \,O_{\mathsf{S}} ]+\frac{C_{\mathsf{B}} }{d_{\mathsf{S}} }\mathrm{Tr}[O_{\mathrm{I}} \,O_{\mathsf{B}} ]\right].
\end{align}
Thus, Eq. \eqref{Ts-connection-2} can be recast as 
\begin{align}
\frac{K_{\mathsf{SB}}}{\widetilde{T}_{\mathsf{SB}}}=\frac{b_{\mathsf{S}}}{\widetilde{T}_{\mathsf{S}}}+\frac{b_{\mathsf{B}}}{\widetilde{T}_{\mathsf{B}}}-\frac{K_{\chi}}{\widetilde{T}_{\chi}}
\label{Ts-connection-3}
\end{align}
by defining the parameters
\begin{align}
K_{\mathsf{SB}}=C_{\mathsf{S}} h_{\mathsf{S}}\left(\frac{\partial U_{\mathsf{SB}}}{\partial U_{\mathsf{S}}}\right)_{u_{2},u_{3},\{r_{i}\}}&=h_{\mathsf{SB}}\left[C_{\mathsf{S}}^{2}+C_{\mathsf{B}}^{2}+C_\chi^{2} \left(\frac{C_{\mathsf{S}}^{2}+C_{\mathsf{B}}^{2}}{C_{\mathsf{S}}^{2} d_{\mathsf{B}} +C_{\mathsf{B}}^{2} d_{\mathsf{S}} }\right) \right],
\label{USB/US}\\
b_{\mathsf{S}}=C_{\mathsf{B}} h_{\mathsf{B}}\left(\frac{\partial U_{\mathsf{S}}}{\partial U_{\mathsf{B}}}\right)_{u_{2},u_{3},\{r_{i}\}}&=C_{\mathsf{S}} h_{\mathsf{S}},
\label{US/US}\\
b_{\mathsf{B}}=C_{\mathsf{S}} h_{\mathsf{S}}\left(\frac{\partial U_{\mathsf{B}}}{\partial U_{\mathsf{S}}}\right)_{u_{2},u_{3},\{r_{i}\}}&=C_{\mathsf{B}} h_{\mathsf{B}},
\label{UB/US}\\
K_{\chi}=C_{\mathsf{S}}h_{\mathsf{S}}\left(\frac{\partial U_{\chi}}{\partial U_{\mathsf{S}}}\right)_{u_{2},u_{3},\{r_{i}\}}&=h_{\mathrm{I}} \left[C_\chi h_{\chi} \left(\frac{C_{\mathsf{S}}^{2}+C_{\mathsf{B}}^{2}}{C_{\mathsf{S}}^{2} d_{\mathsf{B}} +C_{\mathsf{B}}^{2} d_{\mathsf{S}} }\right)+\frac{C_{\mathsf{S}}}{d_{\mathsf{B}}}\mathrm{Tr}[O_{\mathrm{I}} \,O_{\mathsf{S}} ]+\frac{C_{\mathsf{B}} }{d_{\mathsf{S}} }\mathrm{Tr}[O_{\mathrm{I}} \,O_{\mathsf{B}} ]\right].
\end{align}

In the following we consider the special cases of no interaction, no correlation, and large bath and investigate how these limits affect the connection between the temperatures.

\begin{itemize}
\item{\textit{No interaction.}}---When there is no interaction ($H_{\mathrm{I}}=h_{\mathrm{I}} \,O_{\mathrm{I}}=h_{\mathrm{I}}=C_{\chi}=0$). In this case we have from Eq. \eqref{O-sb-expand} that $C_{\mathsf{S}}= h_{\mathsf{S}}/h_{\mathsf{SB}}$ and $C_{\mathsf{B}}= h_{\mathsf{B}}/ h_{\mathsf{SB}}$
which gives
\begin{align}
\frac{h_{\mathsf{S}}^{2}+h_{\mathsf{B}}^{2}}{\widetilde{T}_{\mathsf{SB}}}&=\frac{h_{\mathsf{S}}^{2}}{\widetilde{T}_{\mathsf{S}}}
+ \frac{h_{\mathsf{B}}^{2}}{\widetilde{T}_{\mathsf{B}}}.
\label{no-int}
\end{align}

\item{\textit{No correlation.}}---When there is no correlation between the subsystems, considering that $S_{\chi}=0$ and $U_{\chi}=0$, 
we obtain from Eq. \eqref{Ts-connection-2}
\begin{align}
\frac{K_{\mathsf{SB}}}{\widetilde{T}_{\mathsf{SB}}}=\frac{b_{\mathsf{S}}}{\widetilde{T}_{\mathsf{S}}}+\frac{b_{\mathsf{B}}}{\widetilde{T}_{\mathsf{B}}} .
\label{Ts-connection-no-correlation}
\end{align}

It is important to note that the no-correlation case may occur in two different scenarios: (i) when the correlations vanish at a given instant of time, (ii) when they are always zero. In case (i) we cannot put $S_{\chi}$ and $U_{\chi}$ equal to zero before taking the derivatives. The reason is that $S_{\chi}$ and $U_{\chi}$ change, hence the derivatives of these quantities with respect to other variables, or the derivative of other quantities with respect to them, are not necessarily zero. In case (ii) we can simply put $S_{\chi} = U_{\chi}=0$ everywhere. The derivatives of $S_{\chi}$ and $U_{\chi}$ with respect to anything are zero, and as they are constant in time and do not change, the derivative with respect to them is meaningless. One also should note that case (ii) may not happen in a real physical system, as interactions typically create correlations even when the two subsystems were initially uncorrelated. Based on this remark, Eq. \eqref{Ts-connection-no-correlation} gives the connection between the temperatures at the special instant at which the correlation vanishes. 

\item{\textit{Large bath.}}---If the Hilbert space dimension of the bath is considerably larger than the Hilbert space dimension of the system such that ($d_{\mathsf{B}} \gg d_{\mathsf{S}}$), we obtain the coefficients in Eq. \eqref{Ts-connection-3}. Considering the asymptotic case of $d_{\mathsf{B}}\to \infty$, it is observed from Eq. \eqref{C_S} that $C_{\mathsf{S}}\to 0$. Using this we obtain
\begin{gather}
K_{\mathsf{SB}} = h_{\mathsf{SB}}\left[C_{\mathsf{B}}^{2} + C_\chi^{2} / d_{\mathsf{S}} \right],\\
b_{\mathsf{S}} =0,\\
b_{\mathsf{B}} = h_{\mathsf{B}} C_{\mathsf{B}} ,\\
K_{\chi} = h_{\mathrm{I}}
\left[\frac{C_\chi h_{\chi}}{d_{\mathsf{S}}} + \frac{C_{\mathsf{B}} }{d_{\mathsf{S}}}\mathrm{Tr}[O_{\mathrm{I}} \,O_{\mathsf{B}}]\right].
\end{gather}
By replacing these values in Eq. \eqref{Ts-connection-3} we obtain the connection between temperatures in the large-bath limit. 
\end{itemize}

\section{VII. Obtaining $\widetilde{T}_{\mathsf{S}}$ and $\widetilde{T}_{\mathsf{B}}$ when the global state is Gibbsian}

Assuming that the global state is $\varrho_{\mathsf{SB}}=e^{-\beta H_{\mathsf{SB}}}/Z$, we can obtain $\mathbbmss{H}_{\mathrm{I}}=\beta H_{\mathsf{SB}}+(\log Z) \mathbbmss{I}_{\mathsf{SB}}+\log \varrho_{\mathsf{S}}\otimes \mathbbmss{I}_{\mathsf{B}}+\mathbbmss{I}_{\mathsf{S}} \otimes \log \varrho_{\mathsf{B}}$. From Eq. \eqref{Schi-expanded} in the main text we obtain 
\begin{align}
\left(\frac{\partial S_{\chi}}{\partial U_{\mathsf{S}}}\right)_{U_{\mathsf{B}},U_{\chi},\{r_{i}\}}=&-\frac{1}{2h_{\mathsf{S}}} \mathrm{Tr}[O_{\mathsf{S}}\mathbbmss{H}_{\mathrm{I}}]
+\frac{\mathrm{Tr}[O_{\mathrm{I}}O_{\mathsf{S}}]}{d_{\mathsf{B}}h_{\mathsf{S}}h_{\chi}^{2}}
\left(\mathrm{Tr}[O_{\mathrm{I}}\mathbbmss{H}_{\mathrm{I}}]
-\frac{\mathrm{Tr}[O_{\mathrm{I}}O_{\mathsf{S}}] \, \mathrm{Tr}[O_{\mathsf{S}}\mathbbmss{H}_{\mathrm{I}}]}{d_{\mathsf{B}}}
-\frac{\mathrm{Tr}[O_{\mathrm{I}}O_{\mathsf{B}}] \, \mathrm{Tr}[\mathbbmss{H}_{\mathrm{I}}O_{\mathsf{B}}]}{d_{\mathsf{S}}}
\right)
\nonumber\\
=&-\frac{\beta}{2h_{\mathsf{S}}}\mathrm{Tr}[O_{\mathsf{S}} H_{\mathsf{SB}}] +\frac{1}{h_{\mathsf{S}}}\mathrm{Tr}[O_{\mathsf{S}} \mathbbmss{H}_{\mathsf{S}}]
+\frac{\mathrm{Tr}[O_{\mathrm{I}}O_{\mathsf{S}}]}{d_{\mathsf{B}}h_{\mathsf{S}}h_{\chi}^{2}}
\left(\mathrm{Tr}[O_{\mathrm{I}}\mathbbmss{H}_{\mathrm{I}}]
-\frac{\mathrm{Tr}[O_{\mathrm{I}}O_{\mathsf{S}}] \, \mathrm{Tr}[O_{\mathsf{S}}\mathbbmss{H}_{\mathrm{I}}]}{d_{\mathsf{B}}}
-\frac{\mathrm{Tr}[O_{\mathrm{I}}O_{\mathsf{B}}] \, \mathrm{Tr}[\mathbbmss{H}_{\mathrm{I}}O_{\mathsf{B}}]}{d_{\mathsf{S}}}
\right)
\nonumber\\
=&-\frac{\beta}{2h_{\mathsf{S}}}\mathrm{Tr}[O_{\mathsf{S}} H_{\mathsf{SB}}] +\frac{1}{T_{\mathsf{S}}}
+\frac{\mathrm{Tr}[O_{\mathrm{I}}O_{\mathsf{S}}]}{d_{\mathsf{B}}h_{\mathsf{S}}h_{\chi}^{2}}
\left(\mathrm{Tr}[O_{\mathrm{I}}\mathbbmss{H}_{\mathrm{I}}]
-\frac{\mathrm{Tr}[O_{\mathrm{I}}O_{\mathsf{S}}] \, \mathrm{Tr}[O_{\mathsf{S}}\mathbbmss{H}_{\mathrm{I}}]}{d_{\mathsf{B}}}
-\frac{\mathrm{Tr}[O_{\mathrm{I}}O_{\mathsf{B}}] \, \mathrm{Tr}[\mathbbmss{H}_{\mathrm{I}}O_{\mathsf{B}}]}{d_{\mathsf{S}}}
\right).
\end{align}
Thus, Eq. \eqref{Tglobal-Tlocal} yields
\begin{align}
\frac{1}{\widetilde{T}_{\mathsf{S}}}=\frac{\beta}{2h_{\mathsf{S}}}\mathrm{Tr}[O_{\mathsf{S}} H_{\mathsf{SB}}]
+\frac{\mathrm{Tr}[O_{\mathrm{I}}O_{\mathsf{S}}]}{d_{\mathsf{B}}h_{\mathsf{S}}h_{\chi}^{2}}
\left(\mathrm{Tr}[O_{\mathrm{I}}\mathbbmss{H}_{\mathrm{I}}]
-\frac{\mathrm{Tr}[O_{\mathrm{I}}O_{\mathsf{S}}] \, \mathrm{Tr}[O_{\mathsf{S}}\mathbbmss{H}_{\mathrm{I}}]}{d_{\mathsf{B}}}
-\frac{\mathrm{Tr}[O_{\mathrm{I}}O_{\mathsf{B}}] \, \mathrm{Tr}[\mathbbmss{H}_{\mathrm{I}}O_{\mathsf{B}}]}{d_{\mathsf{S}}}
\right),
\end{align}
 which can be written equivalently as 
\begin{align}
\frac{1}{\widetilde{T}_{\mathsf{S}}}=\beta \Big(1+\frac{1}{2 h_{\mathsf{S}}} \mathrm{Tr}[O_{\mathsf{S}}H_{\mathrm{I}}] \Big)
+\frac{\mathrm{Tr}[O_{\mathrm{I}}O_{\mathsf{S}}]}{d_{\mathsf{B}}h_{\mathsf{S}}h_{\chi}^{2}}
\left(\mathrm{Tr}[O_{\mathrm{I}}\mathbbmss{H}_{\mathrm{I}}] - \frac{\mathrm{Tr}[O_{\mathrm{I}}O_{\mathsf{S}}] \, \mathrm{Tr}[O_{\mathsf{S}}\mathbbmss{H}_{\mathrm{I}}]}{d_{\mathsf{B}}} - \frac{\mathrm{Tr}[O_{\mathrm{I}}O_{\mathsf{B}}] \, \mathrm{Tr}[\mathbbmss{H}_{\mathrm{I}}O_{\mathsf{B}}]}{d_{\mathsf{S}}}
\right),
\end{align}
and similarly for $1/\widetilde{T}_{\mathsf{B}}$. It is observed that when $d_{\mathsf{S}} \ll d_{\mathsf{B}}$, almost all terms on the right-hand side of the above equation which are of the order of $1/d_{\mathsf{B}}$ would have negligible contributions such that $1/\widetilde{T}_{\mathsf{S}} \approx \beta [1 + \mathrm{Tr}[O_{\mathsf{S}}H_{\mathrm{I}}]/(2 h_{\mathsf{S}}) ]$. The correction to $1/\widetilde{T}_{\mathsf{B}}$ becomes small if the coupling ($H_{\mathrm{I}}$) is weak. In addition, by noting that $h_{\mathsf{B}}\approx O(N_{\mathsf{B}})$, with $N_{\mathsf{B}}$ being the number of particles in subsystem $\mathsf{B}$, for $N_{\mathsf{B}} \gg 1$, we obtain $1/\widetilde{T}_{\mathsf{B}} \approx \beta$. This observation agrees with our intuition from standard thermodynamics.



\begin{thebibliography}{83}%
\makeatletter
\providecommand \@ifxundefined [1]{%
 \@ifx{#1\undefined}
}%
\providecommand \@ifnum [1]{%
 \ifnum #1\expandafter \@firstoftwo
 \else \expandafter \@secondoftwo
 \fi
}%
\providecommand \@ifx [1]{%
 \ifx #1\expandafter \@firstoftwo
 \else \expandafter \@secondoftwo
 \fi
}%
\providecommand \natexlab [1]{#1}%
\providecommand \enquote  [1]{``#1''}%
\providecommand \bibnamefont  [1]{#1}%
\providecommand \bibfnamefont [1]{#1}%
\providecommand \citenamefont [1]{#1}%
\providecommand \href@noop [0]{\@secondoftwo}%
\providecommand \href [0]{\begingroup \@sanitize@url \@href}%
\providecommand \@href[1]{\@@startlink{#1}\@@href}%
\providecommand \@@href[1]{\endgroup#1\@@endlink}%
\providecommand \@sanitize@url [0]{\catcode `\\12\catcode `\$12\catcode
  `\&12\catcode `\#12\catcode `\^12\catcode `\_12\catcode `\%12\relax}%
\providecommand \@@startlink[1]{}%
\providecommand \@@endlink[0]{}%
\providecommand \url  [0]{\begingroup\@sanitize@url \@url }%
\providecommand \@url [1]{\endgroup\@href {#1}{\urlprefix }}%
\providecommand \urlprefix  [0]{URL }%
\providecommand \Eprint [0]{\href }%
\providecommand \doibase [0]{https://doi.org/}%
\providecommand \selectlanguage [0]{\@gobble}%
\providecommand \bibinfo  [0]{\@secondoftwo}%
\providecommand \bibfield  [0]{\@secondoftwo}%
\providecommand \translation [1]{[#1]}%
\providecommand \BibitemOpen [0]{}%
\providecommand \bibitemStop [0]{}%
\providecommand \bibitemNoStop [0]{.\EOS\space}%
\providecommand \EOS [0]{\spacefactor3000\relax}%
\providecommand \BibitemShut  [1]{\csname bibitem#1\endcsname}%
\let\auto@bib@innerbib\@empty
\bibitem [{\citenamefont {Pekola}(2015)}]{Pekola-thermo-circuit}%
  \BibitemOpen
  \bibfield  {author} {\bibinfo {author} {\bibfnamefont {J.~P.}\ \bibnamefont
  {Pekola}},\ }\bibfield  {title} {\bibinfo {title} {Towards quantum
  thermodynamics in electronic circuits},\ }\href
  {https://doi.org/10.1038/nphys3169} {\bibfield  {journal} {\bibinfo
  {journal} {Nature Phys.}\ }\textbf {\bibinfo {volume} {11}},\ \bibinfo
  {pages} {118} (\bibinfo {year} {2015})}\BibitemShut {NoStop}%
\bibitem [{\citenamefont {Karimi}\ \emph {et~al.}(2020)\citenamefont {Karimi},
  \citenamefont {Brange}, \citenamefont {Samuelsson},\ and\ \citenamefont
  {Pekola}}]{Pekola-ultimate-resol}%
  \BibitemOpen
  \bibfield  {author} {\bibinfo {author} {\bibfnamefont {B.}~\bibnamefont
  {Karimi}}, \bibinfo {author} {\bibfnamefont {F.}~\bibnamefont {Brange}},
  \bibinfo {author} {\bibfnamefont {P.}~\bibnamefont {Samuelsson}},\ and\
  \bibinfo {author} {\bibfnamefont {J.~P.}\ \bibnamefont {Pekola}},\ }\bibfield
   {title} {\bibinfo {title} {Reaching the ultimate energy resolution of a
  quantum detector},\ }\href {https://doi.org/10.1038/s41467-019-14247-2}
  {\bibfield  {journal} {\bibinfo  {journal} {Nature Commun.}\ }\textbf
  {\bibinfo {volume} {11}},\ \bibinfo {pages} {367} (\bibinfo {year}
  {2020})}\BibitemShut {NoStop}%
\bibitem [{\citenamefont {Kokkoniemi}\ \emph {et~al.}(2020)\citenamefont
  {Kokkoniemi}, \citenamefont {Girard}, \citenamefont {Hazra}, \citenamefont
  {Laitinen}, \citenamefont {Govenius}, \citenamefont {Lake}, \citenamefont
  {Sallinen}, \citenamefont {Vesterinen}, \citenamefont {Partanen},
  \citenamefont {Tan}, \citenamefont {Chan}, \citenamefont {Tan}, \citenamefont
  {Hakonen},\ and\ \citenamefont
  {M\"{o}tt\"{o}nen}}]{Mottonen-Nature-bolometer}%
  \BibitemOpen
  \bibfield  {author} {\bibinfo {author} {\bibfnamefont {R.}~\bibnamefont
  {Kokkoniemi}}, \bibinfo {author} {\bibfnamefont {J.-P.}\ \bibnamefont
  {Girard}}, \bibinfo {author} {\bibfnamefont {D.}~\bibnamefont {Hazra}},
  \bibinfo {author} {\bibfnamefont {A.}~\bibnamefont {Laitinen}}, \bibinfo
  {author} {\bibfnamefont {J.}~\bibnamefont {Govenius}}, \bibinfo {author}
  {\bibfnamefont {R.}~\bibnamefont {Lake}}, \bibinfo {author} {\bibfnamefont
  {I.}~\bibnamefont {Sallinen}}, \bibinfo {author} {\bibfnamefont
  {V.}~\bibnamefont {Vesterinen}}, \bibinfo {author} {\bibfnamefont
  {M.}~\bibnamefont {Partanen}}, \bibinfo {author} {\bibfnamefont
  {J.}~\bibnamefont {Tan}}, \bibinfo {author} {\bibfnamefont {K.~W.}\
  \bibnamefont {Chan}}, \bibinfo {author} {\bibfnamefont {K.~Y.}\ \bibnamefont
  {Tan}}, \bibinfo {author} {\bibfnamefont {P.}~\bibnamefont {Hakonen}},\ and\
  \bibinfo {author} {\bibfnamefont {M.}~\bibnamefont {M\"{o}tt\"{o}nen}},\
  }\bibfield  {title} {\bibinfo {title} {Bolometer operating at the threshold
  for circuit quantum electrodynamics},\ }\href
  {https://doi.org/10.1038/s41586-020-2753-3} {\bibfield  {journal} {\bibinfo
  {journal} {Nature (London)}\ }\textbf {\bibinfo {volume} {586}},\ \bibinfo
  {pages} {47} (\bibinfo {year} {2020})}\BibitemShut {NoStop}%
\bibitem [{\citenamefont {Gemmer}\ \emph {et~al.}(2009)\citenamefont {Gemmer},
  \citenamefont {Michel},\ and\ \citenamefont {Mahler}}]{gemmer2009quantum}%
  \BibitemOpen
  \bibfield  {author} {\bibinfo {author} {\bibfnamefont {J.}~\bibnamefont
  {Gemmer}}, \bibinfo {author} {\bibfnamefont {M.}~\bibnamefont {Michel}},\
  and\ \bibinfo {author} {\bibfnamefont {G.}~\bibnamefont {Mahler}},\
  }\href@noop {} {\emph {\bibinfo {title} {Quantum Thermodynamics: Emergence of
  Thermodynamic Behavior within Composite Quantum Systems}}}\ (\bibinfo
  {publisher} {Springer},\ \bibinfo {address} {Berlin},\ \bibinfo {year}
  {2009})\BibitemShut {NoStop}%
\bibitem [{\citenamefont {Mahler}(2015)}]{book:Malher}%
  \BibitemOpen
  \bibfield  {author} {\bibinfo {author} {\bibfnamefont {G.}~\bibnamefont
  {Mahler}},\ }\href@noop {} {\emph {\bibinfo {title} {Quantum Thermodynamic
  Processes -- Energy and Information Flow at the Nanoscale}}}\ (\bibinfo
  {publisher} {CRC Press},\ \bibinfo {address} {Boca Raton, FL},\ \bibinfo
  {year} {2015})\BibitemShut {NoStop}%
\bibitem [{\citenamefont {Binder}\ \emph {et~al.}(2018)\citenamefont {Binder},
  \citenamefont {Correa}, \citenamefont {Gogolin}, \citenamefont {Anders},\
  and\ \citenamefont {Adesso}}]{book:Qthermo}%
  \BibitemOpen
  \bibinfo {editor} {\bibfnamefont {F.}~\bibnamefont {Binder}}, \bibinfo
  {editor} {\bibfnamefont {L.~A.}\ \bibnamefont {Correa}}, \bibinfo {editor}
  {\bibfnamefont {C.}~\bibnamefont {Gogolin}}, \bibinfo {editor} {\bibfnamefont
  {J.}~\bibnamefont {Anders}},\ and\ \bibinfo {editor} {\bibfnamefont
  {G.}~\bibnamefont {Adesso}},\ eds.,\ \href@noop {} {\emph {\bibinfo {title}
  {Thermodynamics in the Quantum Regime: Fundamental Aspects and New
  Directions}}}\ (\bibinfo  {publisher} {Springer International},\ \bibinfo
  {address} {Cham, Switzerland},\ \bibinfo {year} {2018})\BibitemShut {NoStop}%
\bibitem [{\citenamefont {Deffner}\ and\ \citenamefont
  {Campbell}(2019)}]{book:Deffner-Campbell}%
  \BibitemOpen
  \bibfield  {author} {\bibinfo {author} {\bibfnamefont {S.}~\bibnamefont
  {Deffner}}\ and\ \bibinfo {author} {\bibfnamefont {S.}~\bibnamefont
  {Campbell}},\ }\href@noop {} {\emph {\bibinfo {title} {Quantum Thermodynamics
  -- An Introduction to the Thermodynamics of Quantum Information}}}\ (\bibinfo
   {publisher} {Morgan \& Claypool},\ \bibinfo {address} {San Rafael, CA},\
  \bibinfo {year} {2019})\BibitemShut {NoStop}%
\bibitem [{\citenamefont {Allahverdyan}\ \emph {et~al.}(2004)\citenamefont
  {Allahverdyan}, \citenamefont {Balain},\ and\ \citenamefont
  {Nieuwenhuizen}}]{Allahverdyan--}%
  \BibitemOpen
  \bibfield  {author} {\bibinfo {author} {\bibfnamefont {A.~E.}\ \bibnamefont
  {Allahverdyan}}, \bibinfo {author} {\bibfnamefont {R.}~\bibnamefont
  {Balain}},\ and\ \bibinfo {author} {\bibfnamefont {T.~M.}\ \bibnamefont
  {Nieuwenhuizen}},\ }\bibfield  {title} {\bibinfo {title} {{Quantum
  thermodynamics: Thermodynamics at the nanoscale}},\ }\href
  {https://doi.org/10.1080/09500340408231829} {\bibfield  {journal} {\bibinfo
  {journal} {J. Mod. Opt.}\ }\textbf {\bibinfo {volume} {51}},\ \bibinfo
  {pages} {2703} (\bibinfo {year} {2004})}\BibitemShut {NoStop}%
\bibitem [{\citenamefont {Goldstein}\ \emph {et~al.}(2010)\citenamefont
  {Goldstein}, \citenamefont {Lebowitz}, \citenamefont {Mastrodonato},
  \citenamefont {Tumulka},\ and\ \citenamefont {Zanghi}}]{Goldstein}%
  \BibitemOpen
  \bibfield  {author} {\bibinfo {author} {\bibfnamefont {S.}~\bibnamefont
  {Goldstein}}, \bibinfo {author} {\bibfnamefont {J.~L.}\ \bibnamefont
  {Lebowitz}}, \bibinfo {author} {\bibfnamefont {C.}~\bibnamefont
  {Mastrodonato}}, \bibinfo {author} {\bibfnamefont {R.}~\bibnamefont
  {Tumulka}},\ and\ \bibinfo {author} {\bibfnamefont {N.}~\bibnamefont
  {Zanghi}},\ }\bibfield  {title} {\bibinfo {title} {Approach to thermal
  equilibrium of macroscopic quantum systems},\ }\href
  {https://doi.org/10.1103/PhysRevE.81.011109} {\bibfield  {journal} {\bibinfo
  {journal} {Phys. Rev. E}\ }\textbf {\bibinfo {volume} {81}},\ \bibinfo
  {pages} {011109} (\bibinfo {year} {2010})}\BibitemShut {NoStop}%
\bibitem [{\citenamefont {Goold}\ \emph {et~al.}(2016)\citenamefont {Goold},
  \citenamefont {Huber}, \citenamefont {Riera}, \citenamefont {del Rio},\ and\
  \citenamefont {Skrzypczyk}}]{Goold16}%
  \BibitemOpen
  \bibfield  {author} {\bibinfo {author} {\bibfnamefont {J.}~\bibnamefont
  {Goold}}, \bibinfo {author} {\bibfnamefont {M.}~\bibnamefont {Huber}},
  \bibinfo {author} {\bibfnamefont {A.}~\bibnamefont {Riera}}, \bibinfo
  {author} {\bibfnamefont {L.}~\bibnamefont {del Rio}},\ and\ \bibinfo {author}
  {\bibfnamefont {P.}~\bibnamefont {Skrzypczyk}},\ }\bibfield  {title}
  {\bibinfo {title} {The role of quantum information in thermodynamics -- a
  topical review},\ }\href {https://doi.org/10.1088/1751-8113/49/14/143001}
  {\bibfield  {journal} {\bibinfo  {journal} {J. Phys. A: Math. Theor.}\
  }\textbf {\bibinfo {volume} {49}},\ \bibinfo {pages} {143001} (\bibinfo
  {year} {2016})}\BibitemShut {NoStop}%
\bibitem [{\citenamefont {Gogolin}\ and\ \citenamefont
  {Eisert}(2016)}]{Gogolin-Eisert}%
  \BibitemOpen
  \bibfield  {author} {\bibinfo {author} {\bibfnamefont {C.}~\bibnamefont
  {Gogolin}}\ and\ \bibinfo {author} {\bibfnamefont {J.}~\bibnamefont
  {Eisert}},\ }\bibfield  {title} {\bibinfo {title} {Equilibration,
  thermalisation, and the emergence of statistical mechanics in closed quantum
  systems},\ }\href {https://doi.org/10.1088/0034-4885/79/5/056001} {\bibfield
  {journal} {\bibinfo  {journal} {Rep. Prog. Phys.}\ }\textbf {\bibinfo
  {volume} {79}},\ \bibinfo {pages} {056001} (\bibinfo {year}
  {2016})}\BibitemShut {NoStop}%
\bibitem [{\citenamefont {Allahverdyan}\ and\ \citenamefont
  {Nieuwenhuizen}(2000)}]{Allahverdyan-}%
  \BibitemOpen
  \bibfield  {author} {\bibinfo {author} {\bibfnamefont {A.~E.}\ \bibnamefont
  {Allahverdyan}}\ and\ \bibinfo {author} {\bibfnamefont {T.~M.}\ \bibnamefont
  {Nieuwenhuizen}},\ }\bibfield  {title} {\bibinfo {title} {{Extraction of Work
  from a Single Thermal Bath in the Quantum Regime}},\ }\href
  {https://doi.org/10.1103/PhysRevLett.85.1799} {\bibfield  {journal} {\bibinfo
   {journal} {Phys. Rev. Lett.}\ }\textbf {\bibinfo {volume} {85}},\ \bibinfo
  {pages} {1799} (\bibinfo {year} {2000})}\BibitemShut {NoStop}%
\bibitem [{\citenamefont {Scully}\ \emph {et~al.}(2003)\citenamefont {Scully},
  \citenamefont {Zubairy}, \citenamefont {Agarwal},\ and\ \citenamefont
  {Walther}}]{Scully-SingleHeatBath}%
  \BibitemOpen
  \bibfield  {author} {\bibinfo {author} {\bibfnamefont {M.~O.}\ \bibnamefont
  {Scully}}, \bibinfo {author} {\bibfnamefont {M.~S.}\ \bibnamefont {Zubairy}},
  \bibinfo {author} {\bibfnamefont {G.~S.}\ \bibnamefont {Agarwal}},\ and\
  \bibinfo {author} {\bibfnamefont {H.}~\bibnamefont {Walther}},\ }\bibfield
  {title} {\bibinfo {title} {Extracting work from a single heat bath via
  vanishing quantum coherence},\ }\href
  {https://doi.org/10.1126/science.1078955} {\bibfield  {journal} {\bibinfo
  {journal} {Science}\ }\textbf {\bibinfo {volume} {299}},\ \bibinfo {pages}
  {862} (\bibinfo {year} {2003})}\BibitemShut {NoStop}%
\bibitem [{\citenamefont {Alipour}\ \emph {et~al.}(2016)\citenamefont
  {Alipour}, \citenamefont {Benatti}, \citenamefont {Bakhshinezhad},
  \citenamefont {Afsary}, \citenamefont {Marcantoni},\ and\ \citenamefont
  {Rezakhani}}]{SciRep}%
  \BibitemOpen
  \bibfield  {author} {\bibinfo {author} {\bibfnamefont {S.}~\bibnamefont
  {Alipour}}, \bibinfo {author} {\bibfnamefont {F.}~\bibnamefont {Benatti}},
  \bibinfo {author} {\bibfnamefont {F.}~\bibnamefont {Bakhshinezhad}}, \bibinfo
  {author} {\bibfnamefont {M.}~\bibnamefont {Afsary}}, \bibinfo {author}
  {\bibfnamefont {S.}~\bibnamefont {Marcantoni}},\ and\ \bibinfo {author}
  {\bibfnamefont {A.~T.}\ \bibnamefont {Rezakhani}},\ }\bibfield  {title}
  {\bibinfo {title} {Correlations in quantum thermodynamics: Heat, work, and
  entropy production},\ }\href {https://doi.org/10.1038/srep35568} {\bibfield
  {journal} {\bibinfo  {journal} {Sci. Rep.}\ }\textbf {\bibinfo {volume}
  {6}},\ \bibinfo {pages} {35568} (\bibinfo {year} {2016})}\BibitemShut
  {NoStop}%
\bibitem [{\citenamefont {Vinjanampathy}\ and\ \citenamefont
  {Modi}(2016)}]{Vinjanampathy-Modi-initialCorr}%
  \BibitemOpen
  \bibfield  {author} {\bibinfo {author} {\bibfnamefont {S.}~\bibnamefont
  {Vinjanampathy}}\ and\ \bibinfo {author} {\bibfnamefont {K.}~\bibnamefont
  {Modi}},\ }\bibfield  {title} {\bibinfo {title} {Correlations, operations and
  the second law of thermodynamics},\ }\href
  {https://doi.org/10.1142/S0219749916400335} {\bibfield  {journal} {\bibinfo
  {journal} {Int. J. Quantum Inf.}\ }\textbf {\bibinfo {volume} {14}},\
  \bibinfo {pages} {1640033} (\bibinfo {year} {2016})}\BibitemShut {NoStop}%
\bibitem [{\citenamefont {Bera}\ \emph {et~al.}(2017)\citenamefont {Bera},
  \citenamefont {Riera}, \citenamefont {Lewenstein},\ and\ \citenamefont
  {Winter}}]{bera-GeneralizedLawsThermo}%
  \BibitemOpen
  \bibfield  {author} {\bibinfo {author} {\bibfnamefont {M.~N.}\ \bibnamefont
  {Bera}}, \bibinfo {author} {\bibfnamefont {A.}~\bibnamefont {Riera}},
  \bibinfo {author} {\bibfnamefont {M.}~\bibnamefont {Lewenstein}},\ and\
  \bibinfo {author} {\bibfnamefont {A.}~\bibnamefont {Winter}},\ }\bibfield
  {title} {\bibinfo {title} {Generalized laws of thermodynamics in the presence
  of correlations},\ }\href {https://doi.org/10.1038/s41467-017-02370-x}
  {\bibfield  {journal} {\bibinfo  {journal} {Nature Commun.}\ }\textbf
  {\bibinfo {volume} {8}},\ \bibinfo {pages} {2180} (\bibinfo {year}
  {2017})}\BibitemShut {NoStop}%
\bibitem [{\citenamefont {Su}\ \emph {et~al.}(2018)\citenamefont {Su},
  \citenamefont {Chen}, \citenamefont {Ma}, \citenamefont {Chen},\ and\
  \citenamefont {Sun}}]{Su-heatCoherence}%
  \BibitemOpen
  \bibfield  {author} {\bibinfo {author} {\bibfnamefont {S.}~\bibnamefont
  {Su}}, \bibinfo {author} {\bibfnamefont {J.}~\bibnamefont {Chen}}, \bibinfo
  {author} {\bibfnamefont {Y.}~\bibnamefont {Ma}}, \bibinfo {author}
  {\bibfnamefont {J.}~\bibnamefont {Chen}},\ and\ \bibinfo {author}
  {\bibfnamefont {C.}~\bibnamefont {Sun}},\ }\bibfield  {title} {\bibinfo
  {title} {The heat and work of quantum thermodynamic processes with quantum
  coherence},\ }\href {https://doi.org/10.1088/1674-1056/27/6/060502}
  {\bibfield  {journal} {\bibinfo  {journal} {Chin. Phys. B}\ }\textbf
  {\bibinfo {volume} {27}},\ \bibinfo {pages} {060502} (\bibinfo {year}
  {2018})}\BibitemShut {NoStop}%
\bibitem [{\citenamefont {Bakhshinezhad}\ \emph {et~al.}(2019)\citenamefont
  {Bakhshinezhad}, \citenamefont {Clivaz}, \citenamefont {Vitagliano},
  \citenamefont {Erker}, \citenamefont {Rezakhani}, \citenamefont {Huber},\
  and\ \citenamefont {Friis}}]{Bakhshinezhad_2019}%
  \BibitemOpen
  \bibfield  {author} {\bibinfo {author} {\bibfnamefont {F.}~\bibnamefont
  {Bakhshinezhad}}, \bibinfo {author} {\bibfnamefont {F.}~\bibnamefont
  {Clivaz}}, \bibinfo {author} {\bibfnamefont {G.}~\bibnamefont {Vitagliano}},
  \bibinfo {author} {\bibfnamefont {P.}~\bibnamefont {Erker}}, \bibinfo
  {author} {\bibfnamefont {A.}~\bibnamefont {Rezakhani}}, \bibinfo {author}
  {\bibfnamefont {M.}~\bibnamefont {Huber}},\ and\ \bibinfo {author}
  {\bibfnamefont {N.}~\bibnamefont {Friis}},\ }\bibfield  {title} {\bibinfo
  {title} {Thermodynamically optimal creation of correlations},\ }\href
  {https://doi.org/10.1088/1751-8121/ab3932} {\bibfield  {journal} {\bibinfo
  {journal} {J. Phys. A: Math. Theor.}\ }\textbf {\bibinfo {volume} {52}},\
  \bibinfo {pages} {465303} (\bibinfo {year} {2019})}\BibitemShut {NoStop}%
\bibitem [{\citenamefont {Manzano}\ \emph {et~al.}(2018)\citenamefont
  {Manzano}, \citenamefont {Plastina},\ and\ \citenamefont
  {Zambrini}}]{Manzano-work-corr}%
  \BibitemOpen
  \bibfield  {author} {\bibinfo {author} {\bibfnamefont {G.}~\bibnamefont
  {Manzano}}, \bibinfo {author} {\bibfnamefont {F.}~\bibnamefont {Plastina}},\
  and\ \bibinfo {author} {\bibfnamefont {R.}~\bibnamefont {Zambrini}},\
  }\bibfield  {title} {\bibinfo {title} {{Optimal Work Extraction and
  Thermodynamics of Quantum Measurements and Correlations}},\ }\href
  {https://doi.org/10.1103/PhysRevLett.121.120602} {\bibfield  {journal}
  {\bibinfo  {journal} {Phys. Rev. Lett.}\ }\textbf {\bibinfo {volume} {121}},\
  \bibinfo {pages} {120602} (\bibinfo {year} {2018})}\BibitemShut {NoStop}%
\bibitem [{\citenamefont {Micadei}\ \emph {et~al.}(2019)\citenamefont
  {Micadei}, \citenamefont {Peterson}, \citenamefont {Souza}, \citenamefont
  {Sarthour}, \citenamefont {Oliveira}, \citenamefont {Landi}, \citenamefont
  {Batalh{\~a}o}, \citenamefont {Serra},\ and\ \citenamefont
  {Lutz}}]{Lutz-Reverse-Heat-flow}%
  \BibitemOpen
  \bibfield  {author} {\bibinfo {author} {\bibfnamefont {K.}~\bibnamefont
  {Micadei}}, \bibinfo {author} {\bibfnamefont {J.~P.}\ \bibnamefont
  {Peterson}}, \bibinfo {author} {\bibfnamefont {A.~M.}\ \bibnamefont {Souza}},
  \bibinfo {author} {\bibfnamefont {R.~S.}\ \bibnamefont {Sarthour}}, \bibinfo
  {author} {\bibfnamefont {I.~S.}\ \bibnamefont {Oliveira}}, \bibinfo {author}
  {\bibfnamefont {G.~T.}\ \bibnamefont {Landi}}, \bibinfo {author}
  {\bibfnamefont {T.~B.}\ \bibnamefont {Batalh{\~a}o}}, \bibinfo {author}
  {\bibfnamefont {R.~M.}\ \bibnamefont {Serra}},\ and\ \bibinfo {author}
  {\bibfnamefont {E.}~\bibnamefont {Lutz}},\ }\bibfield  {title} {\bibinfo
  {title} {Reversing the direction of heat flow using quantum correlations},\
  }\href {https://doi.org/10.1038/s41467-019-10333-7} {\bibfield  {journal}
  {\bibinfo  {journal} {Nature Commun.}\ }\textbf {\bibinfo {volume} {10}},\
  \bibinfo {pages} {2456} (\bibinfo {year} {2019})}\BibitemShut {NoStop}%
\bibitem [{\citenamefont {Alipour}\ \emph
  {et~al.}(2020{\natexlab{a}})\citenamefont {Alipour}, \citenamefont {Tuohino},
  \citenamefont {Rezakhani},\ and\ \citenamefont
  {Ala-Nissila}}]{unreliability-mutual-info}%
  \BibitemOpen
  \bibfield  {author} {\bibinfo {author} {\bibfnamefont {S.}~\bibnamefont
  {Alipour}}, \bibinfo {author} {\bibfnamefont {S.}~\bibnamefont {Tuohino}},
  \bibinfo {author} {\bibfnamefont {A.~T.}\ \bibnamefont {Rezakhani}},\ and\
  \bibinfo {author} {\bibfnamefont {T.}~\bibnamefont {Ala-Nissila}},\
  }\bibfield  {title} {\bibinfo {title} {Unreliability of mutual information as
  a measure for variations in total correlations},\ }\href
  {https://doi.org/10.1103/PhysRevA.101.042311} {\bibfield  {journal} {\bibinfo
   {journal} {Phys. Rev. A}\ }\textbf {\bibinfo {volume} {101}},\ \bibinfo
  {pages} {042311} (\bibinfo {year} {2020}{\natexlab{a}})}\BibitemShut
  {NoStop}%
\bibitem [{\citenamefont {Kammerlander}\ and\ \citenamefont
  {Anders}(2016)}]{Anders-coherence}%
  \BibitemOpen
  \bibfield  {author} {\bibinfo {author} {\bibfnamefont {P.}~\bibnamefont
  {Kammerlander}}\ and\ \bibinfo {author} {\bibfnamefont {J.}~\bibnamefont
  {Anders}},\ }\bibfield  {title} {\bibinfo {title} {Coherence and measurement
  in quantum thermodynamics},\ }\href {https://doi.org/10.1038/srep22174}
  {\bibfield  {journal} {\bibinfo  {journal} {Sci. Rep.}\ }\textbf {\bibinfo
  {volume} {6}},\ \bibinfo {pages} {22174} (\bibinfo {year}
  {2016})}\BibitemShut {NoStop}%
\bibitem [{\citenamefont {D{\'\i}az}\ \emph {et~al.}(2020)\citenamefont
  {D{\'\i}az}, \citenamefont {Guarnieri},\ and\ \citenamefont
  {Paternostro}}]{Paternostro-work-initial-coherence}%
  \BibitemOpen
  \bibfield  {author} {\bibinfo {author} {\bibfnamefont {M.~G.}\ \bibnamefont
  {D{\'\i}az}}, \bibinfo {author} {\bibfnamefont {G.}~\bibnamefont
  {Guarnieri}},\ and\ \bibinfo {author} {\bibfnamefont {M.}~\bibnamefont
  {Paternostro}},\ }\bibfield  {title} {\bibinfo {title} {Quantum work
  statistics with initial coherence},\ }\href
  {https://doi.org/10.3390/e22111223} {\bibfield  {journal} {\bibinfo
  {journal} {Entropy}\ }\textbf {\bibinfo {volume} {22}},\ \bibinfo {pages}
  {1223} (\bibinfo {year} {2020})}\BibitemShut {NoStop}%
\bibitem [{\citenamefont {Marcantoni}\ \emph {et~al.}(2017)\citenamefont
  {Marcantoni}, \citenamefont {Alipour}, \citenamefont {Benatti}, \citenamefont
  {Floreanini},\ and\ \citenamefont {Rezakhani}}]{SciRep-entropy-production}%
  \BibitemOpen
  \bibfield  {author} {\bibinfo {author} {\bibfnamefont {S.}~\bibnamefont
  {Marcantoni}}, \bibinfo {author} {\bibfnamefont {S.}~\bibnamefont {Alipour}},
  \bibinfo {author} {\bibfnamefont {F.}~\bibnamefont {Benatti}}, \bibinfo
  {author} {\bibfnamefont {R.}~\bibnamefont {Floreanini}},\ and\ \bibinfo
  {author} {\bibfnamefont {A.~T.}\ \bibnamefont {Rezakhani}},\ }\bibfield
  {title} {\bibinfo {title} {{Entropy production and non-Markovian dynamical
  maps}},\ }\href {https://doi.org/10.1038/s41598-017-12595-x} {\bibfield
  {journal} {\bibinfo  {journal} {Sci. Rep.}\ }\textbf {\bibinfo {volume}
  {7}},\ \bibinfo {pages} {12447} (\bibinfo {year} {2017})}\BibitemShut
  {NoStop}%
\bibitem [{\citenamefont {Strasberg}\ and\ \citenamefont
  {Esposito}(2019)}]{Strasberg-entroproduct}%
  \BibitemOpen
  \bibfield  {author} {\bibinfo {author} {\bibfnamefont {P.}~\bibnamefont
  {Strasberg}}\ and\ \bibinfo {author} {\bibfnamefont {M.}~\bibnamefont
  {Esposito}},\ }\bibfield  {title} {\bibinfo {title} {{Non-Markovianity} and
  negative entropy production rates},\ }\href
  {https://doi.org/10.1103/PhysRevE.99.012120} {\bibfield  {journal} {\bibinfo
  {journal} {Phys. Rev. E}\ }\textbf {\bibinfo {volume} {99}},\ \bibinfo
  {pages} {012120} (\bibinfo {year} {2019})}\BibitemShut {NoStop}%
\bibitem [{\citenamefont {Manzano}\ \emph {et~al.}(2019)\citenamefont
  {Manzano}, \citenamefont {Fazio},\ and\ \citenamefont
  {Rold\'an}}]{Manzano-martingale-entropyProduction}%
  \BibitemOpen
  \bibfield  {author} {\bibinfo {author} {\bibfnamefont {G.}~\bibnamefont
  {Manzano}}, \bibinfo {author} {\bibfnamefont {R.}~\bibnamefont {Fazio}},\
  and\ \bibinfo {author} {\bibfnamefont {E.}~\bibnamefont {Rold\'an}},\
  }\bibfield  {title} {\bibinfo {title} {{Quantum Martingale Theory and Entropy
  Production}},\ }\href {https://doi.org/10.1103/PhysRevLett.122.220602}
  {\bibfield  {journal} {\bibinfo  {journal} {Phys. Rev. Lett.}\ }\textbf
  {\bibinfo {volume} {122}},\ \bibinfo {pages} {220602} (\bibinfo {year}
  {2019})}\BibitemShut {NoStop}%
\bibitem [{\citenamefont {Esposito}\ \emph {et~al.}(2009)\citenamefont
  {Esposito}, \citenamefont {Harbola},\ and\ \citenamefont
  {Mukamel}}]{Esposito09}%
  \BibitemOpen
  \bibfield  {author} {\bibinfo {author} {\bibfnamefont {M.}~\bibnamefont
  {Esposito}}, \bibinfo {author} {\bibfnamefont {U.}~\bibnamefont {Harbola}},\
  and\ \bibinfo {author} {\bibfnamefont {S.}~\bibnamefont {Mukamel}},\
  }\bibfield  {title} {\bibinfo {title} {Nonequilibrium fluctuations,
  fluctuation theorems, and counting statistics in quantum systems},\ }\href
  {https://doi.org/10.1103/RevModPhys.81.1665} {\bibfield  {journal} {\bibinfo
  {journal} {Rev. Mod. Phys.}\ }\textbf {\bibinfo {volume} {81}},\ \bibinfo
  {pages} {1665} (\bibinfo {year} {2009})}\BibitemShut {NoStop}%
\bibitem [{\citenamefont {Campisi}\ \emph {et~al.}(2011)\citenamefont
  {Campisi}, \citenamefont {H\"anggi},\ and\ \citenamefont
  {Talkner}}]{Campisi11}%
  \BibitemOpen
  \bibfield  {author} {\bibinfo {author} {\bibfnamefont {M.}~\bibnamefont
  {Campisi}}, \bibinfo {author} {\bibfnamefont {P.}~\bibnamefont {H\"anggi}},\
  and\ \bibinfo {author} {\bibfnamefont {P.}~\bibnamefont {Talkner}},\
  }\bibfield  {title} {\bibinfo {title} {Quantum fluctuation relations:
  Foundations and applications},\ }\href
  {https://doi.org/10.1103/RevModPhys.83.771} {\bibfield  {journal} {\bibinfo
  {journal} {Rev. Mod. Phys.}\ }\textbf {\bibinfo {volume} {83}},\ \bibinfo
  {pages} {771} (\bibinfo {year} {2011})}\BibitemShut {NoStop}%
\bibitem [{\citenamefont {Funo}\ \emph {et~al.}(2018)\citenamefont {Funo},
  \citenamefont {Ueda},\ and\ \citenamefont {Sagawa}}]{Funo18}%
  \BibitemOpen
  \bibfield  {author} {\bibinfo {author} {\bibfnamefont {K.}~\bibnamefont
  {Funo}}, \bibinfo {author} {\bibfnamefont {M.}~\bibnamefont {Ueda}},\ and\
  \bibinfo {author} {\bibfnamefont {T.}~\bibnamefont {Sagawa}},\ }\bibinfo
  {title} {Quantum fluctuation theorems},\ in\ \href@noop {} {\emph {\bibinfo
  {booktitle} {Thermodynamics in the Quantum Regime: Fundamental Aspects and
  New Directions}}},\ \bibinfo {editor} {edited by\ \bibinfo {editor}
  {\bibfnamefont {F.}~\bibnamefont {Binder}}, \bibinfo {editor} {\bibfnamefont
  {L.~A.}\ \bibnamefont {Correa}}, \bibinfo {editor} {\bibfnamefont
  {C.}~\bibnamefont {Gogolin}}, \bibinfo {editor} {\bibfnamefont
  {J.}~\bibnamefont {Anders}},\ and\ \bibinfo {editor} {\bibfnamefont
  {G.}~\bibnamefont {Adesso}}}\ (\bibinfo  {publisher} {Springer
  International},\ \bibinfo {address} {Cham, Switzerland},\ \bibinfo {year}
  {2018})\ p.\ \bibinfo {pages} {249}\BibitemShut {NoStop}%
\bibitem [{\citenamefont {Ramezani}\ \emph {et~al.}(2018)\citenamefont
  {Ramezani}, \citenamefont {Benatti}, \citenamefont {Floreanini},
  \citenamefont {Marcantoni}, \citenamefont {Golshani},\ and\ \citenamefont
  {Rezakhani}}]{ATR-qft2}%
  \BibitemOpen
  \bibfield  {author} {\bibinfo {author} {\bibfnamefont {M.}~\bibnamefont
  {Ramezani}}, \bibinfo {author} {\bibfnamefont {F.}~\bibnamefont {Benatti}},
  \bibinfo {author} {\bibfnamefont {R.}~\bibnamefont {Floreanini}}, \bibinfo
  {author} {\bibfnamefont {S.}~\bibnamefont {Marcantoni}}, \bibinfo {author}
  {\bibfnamefont {M.}~\bibnamefont {Golshani}},\ and\ \bibinfo {author}
  {\bibfnamefont {A.~T.}\ \bibnamefont {Rezakhani}},\ }\bibfield  {title}
  {\bibinfo {title} {Quantum detailed balance conditions and fluctuation
  relations for thermalizing quantum dynamics},\ }\href
  {https://doi.org/10.1103/PhysRevE.98.052104} {\bibfield  {journal} {\bibinfo
  {journal} {Phys. Rev. E}\ }\textbf {\bibinfo {volume} {98}},\ \bibinfo
  {pages} {052104} (\bibinfo {year} {2018})}\BibitemShut {NoStop}%
\bibitem [{\citenamefont {Garc\'{\i}a-Pintos}\ \emph
  {et~al.}(2020)\citenamefont {Garc\'{\i}a-Pintos}, \citenamefont {Hamma},\
  and\ \citenamefont {del Campo}}]{delCampo-fluctuation-battery}%
  \BibitemOpen
  \bibfield  {author} {\bibinfo {author} {\bibfnamefont {L.~P.}\ \bibnamefont
  {Garc\'{\i}a-Pintos}}, \bibinfo {author} {\bibfnamefont {A.}~\bibnamefont
  {Hamma}},\ and\ \bibinfo {author} {\bibfnamefont {A.}~\bibnamefont {del
  Campo}},\ }\bibfield  {title} {\bibinfo {title} {{Fluctuations in Extractable
  Work Bound the Charging Power of Quantum Batteries}},\ }\href
  {https://doi.org/10.1103/PhysRevLett.125.040601} {\bibfield  {journal}
  {\bibinfo  {journal} {Phys. Rev. Lett.}\ }\textbf {\bibinfo {volume} {125}},\
  \bibinfo {pages} {040601} (\bibinfo {year} {2020})}\BibitemShut {NoStop}%
\bibitem [{\citenamefont {Hilt}\ \emph {et~al.}(2011)\citenamefont {Hilt},
  \citenamefont {Shabbir}, \citenamefont {Anders},\ and\ \citenamefont
  {Lutz}}]{Anders-Lutz-Landauer}%
  \BibitemOpen
  \bibfield  {author} {\bibinfo {author} {\bibfnamefont {S.}~\bibnamefont
  {Hilt}}, \bibinfo {author} {\bibfnamefont {S.}~\bibnamefont {Shabbir}},
  \bibinfo {author} {\bibfnamefont {J.}~\bibnamefont {Anders}},\ and\ \bibinfo
  {author} {\bibfnamefont {E.}~\bibnamefont {Lutz}},\ }\bibfield  {title}
  {\bibinfo {title} {Landauer's principle in the quantum regime},\ }\href
  {https://doi.org/10.1103/PhysRevE.83.030102} {\bibfield  {journal} {\bibinfo
  {journal} {Phys. Rev. E}\ }\textbf {\bibinfo {volume} {83}},\ \bibinfo
  {pages} {030102} (\bibinfo {year} {2011})}\BibitemShut {NoStop}%
\bibitem [{\citenamefont {Kosloff}(2013)}]{Kosloff13}%
  \BibitemOpen
  \bibfield  {author} {\bibinfo {author} {\bibfnamefont {R.}~\bibnamefont
  {Kosloff}},\ }\bibfield  {title} {\bibinfo {title} {Quantum thermodynamics: A
  dynamical viewpoint},\ }\href {https://doi.org/10.3390/e15062100} {\bibfield
  {journal} {\bibinfo  {journal} {Entropy}\ }\textbf {\bibinfo {volume} {15}},\
  \bibinfo {pages} {2100} (\bibinfo {year} {2013})}\BibitemShut {NoStop}%
\bibitem [{\citenamefont {del Campo}\ \emph {et~al.}(2018)\citenamefont {del
  Campo}, \citenamefont {Chenu}, \citenamefont {Deng},\ and\ \citenamefont
  {Wu}}]{delcampo18}%
  \BibitemOpen
  \bibfield  {author} {\bibinfo {author} {\bibfnamefont {A.}~\bibnamefont {del
  Campo}}, \bibinfo {author} {\bibfnamefont {A.}~\bibnamefont {Chenu}},
  \bibinfo {author} {\bibfnamefont {S.}~\bibnamefont {Deng}},\ and\ \bibinfo
  {author} {\bibfnamefont {H.}~\bibnamefont {Wu}},\ }\bibinfo {title}
  {Friction-free quantum machines},\ in\ \href@noop {} {\emph {\bibinfo
  {booktitle} {Thermodynamics in the Quantum Regime: Fundamental Aspects and
  New Directions}}},\ \bibinfo {editor} {edited by\ \bibinfo {editor}
  {\bibfnamefont {F.}~\bibnamefont {Binder}}, \bibinfo {editor} {\bibfnamefont
  {L.~A.}\ \bibnamefont {Correa}}, \bibinfo {editor} {\bibfnamefont
  {C.}~\bibnamefont {Gogolin}}, \bibinfo {editor} {\bibfnamefont
  {J.}~\bibnamefont {Anders}},\ and\ \bibinfo {editor} {\bibfnamefont
  {G.}~\bibnamefont {Adesso}}}\ (\bibinfo  {publisher} {Springer
  International},\ \bibinfo {address} {Cham, Switzerland},\ \bibinfo {year}
  {2018})\ p.\ \bibinfo {pages} {127}\BibitemShut {NoStop}%
\bibitem [{\citenamefont {Rivas}(2020)}]{Rivas-nonEquilibMeanForce}%
  \BibitemOpen
  \bibfield  {author} {\bibinfo {author} {\bibfnamefont {A.}~\bibnamefont
  {Rivas}},\ }\bibfield  {title} {\bibinfo {title} {{Strong Coupling
  Thermodynamics of Open Quantum Systems}},\ }\href
  {https://doi.org/10.1103/PhysRevLett.124.160601} {\bibfield  {journal}
  {\bibinfo  {journal} {Phys. Rev. Lett.}\ }\textbf {\bibinfo {volume} {124}},\
  \bibinfo {pages} {160601} (\bibinfo {year} {2020})}\BibitemShut {NoStop}%
\bibitem [{\citenamefont {Ramezani}\ \emph {et~al.}(2019)\citenamefont
  {Ramezani}, \citenamefont {Marcantoni}, \citenamefont {Benatti},
  \citenamefont {Floreanini}, \citenamefont {Petiziol}, \citenamefont
  {Rezakhani},\ and\ \citenamefont {Golshani}}]{ATR-carnot}%
  \BibitemOpen
  \bibfield  {author} {\bibinfo {author} {\bibfnamefont {M.}~\bibnamefont
  {Ramezani}}, \bibinfo {author} {\bibfnamefont {S.}~\bibnamefont
  {Marcantoni}}, \bibinfo {author} {\bibfnamefont {F.}~\bibnamefont {Benatti}},
  \bibinfo {author} {\bibfnamefont {R.}~\bibnamefont {Floreanini}}, \bibinfo
  {author} {\bibfnamefont {F.}~\bibnamefont {Petiziol}}, \bibinfo {author}
  {\bibfnamefont {A.~T.}\ \bibnamefont {Rezakhani}},\ and\ \bibinfo {author}
  {\bibfnamefont {M.}~\bibnamefont {Golshani}},\ }\bibfield  {title} {\bibinfo
  {title} {Impact of nonideal cycles on the efficiency of quantum heat
  engines},\ }\href {https://doi.org/10.1140/epjd/e2019-90520-7} {\bibfield
  {journal} {\bibinfo  {journal} {Eur. Phys. J. D}\ }\textbf {\bibinfo {volume}
  {73}},\ \bibinfo {pages} {144} (\bibinfo {year} {2019})}\BibitemShut
  {NoStop}%
\bibitem [{\citenamefont {Alicki}(1979)}]{Alicki}%
  \BibitemOpen
  \bibfield  {author} {\bibinfo {author} {\bibfnamefont {R.}~\bibnamefont
  {Alicki}},\ }\bibfield  {title} {\bibinfo {title} {The quantum open system as
  a model of the heat engine},\ }\href
  {https://doi.org/10.1088/0305-4470/12/5/007} {\bibfield  {journal} {\bibinfo
  {journal} {J. Phys. A: Math. Gen.}\ }\textbf {\bibinfo {volume} {12}},\
  \bibinfo {pages} {L103} (\bibinfo {year} {1979})}\BibitemShut {NoStop}%
\bibitem [{\citenamefont {Spohn}(1978)}]{Spohn-EP}%
  \BibitemOpen
  \bibfield  {author} {\bibinfo {author} {\bibfnamefont {H.}~\bibnamefont
  {Spohn}},\ }\bibfield  {title} {\bibinfo {title} {Entropy production for
  quantum dynamical semigroups},\ }\href {https://doi.org/10.1063/1.523789}
  {\bibfield  {journal} {\bibinfo  {journal} {J. Math. Phys.}\ }\textbf
  {\bibinfo {volume} {19}},\ \bibinfo {pages} {1227} (\bibinfo {year}
  {1978})}\BibitemShut {NoStop}%
\bibitem [{\citenamefont {Niedenzu}\ \emph {et~al.}(2018)\citenamefont
  {Niedenzu}, \citenamefont {Mukherjee}, \citenamefont {Ghosh}, \citenamefont
  {Kofman},\ and\ \citenamefont {Kurizki}}]{Kurizki-ergo}%
  \BibitemOpen
  \bibfield  {author} {\bibinfo {author} {\bibfnamefont {W.}~\bibnamefont
  {Niedenzu}}, \bibinfo {author} {\bibfnamefont {V.}~\bibnamefont {Mukherjee}},
  \bibinfo {author} {\bibfnamefont {A.}~\bibnamefont {Ghosh}}, \bibinfo
  {author} {\bibfnamefont {A.~G.}\ \bibnamefont {Kofman}},\ and\ \bibinfo
  {author} {\bibfnamefont {G.}~\bibnamefont {Kurizki}},\ }\bibfield  {title}
  {\bibinfo {title} {Quantum engine efficiency bound beyond the second law of
  thermodynamics},\ }\href {https://doi.org/10.1038/s41467-017-01991-6}
  {\bibfield  {journal} {\bibinfo  {journal} {Nature Commun.}\ }\textbf
  {\bibinfo {volume} {9}},\ \bibinfo {pages} {165} (\bibinfo {year}
  {2018})}\BibitemShut {NoStop}%
\bibitem [{\citenamefont {Polkovnikov}(2011)}]{Polkovnikov-diag-entr}%
  \BibitemOpen
  \bibfield  {author} {\bibinfo {author} {\bibfnamefont {A.}~\bibnamefont
  {Polkovnikov}},\ }\bibfield  {title} {\bibinfo {title} {Microscopic diagonal
  entropy and its connection to basic thermodynamic relations},\ }\href
  {https://doi.org/10.1016/j.aop.2010.08.004} {\bibfield  {journal} {\bibinfo
  {journal} {Ann. Phys. (N.Y.)}\ }\textbf {\bibinfo {volume} {326}},\ \bibinfo
  {pages} {486} (\bibinfo {year} {2011})}\BibitemShut {NoStop}%
\bibitem [{\citenamefont {Alipour}\ \emph {et~al.}(2019)\citenamefont
  {Alipour}, \citenamefont {Rezakhani}, \citenamefont {Chenu}, \citenamefont
  {del Campo},\ and\ \citenamefont {Ala-Nissila}}]{entropic-based-heat-work}%
  \BibitemOpen
  \bibfield  {author} {\bibinfo {author} {\bibfnamefont {S.}~\bibnamefont
  {Alipour}}, \bibinfo {author} {\bibfnamefont {A.~T.}\ \bibnamefont
  {Rezakhani}}, \bibinfo {author} {\bibfnamefont {A.}~\bibnamefont {Chenu}},
  \bibinfo {author} {\bibfnamefont {A.}~\bibnamefont {del Campo}},\ and\
  \bibinfo {author} {\bibfnamefont {T.}~\bibnamefont {Ala-Nissila}},\
  }\bibfield  {title} {\bibinfo {title} {Unambiguous formulation for heat and
  work in arbitrary quantum evolution},\ }\href
  {https://arxiv.org/abs/1912.01939} {\bibfield  {journal} {\bibinfo  {journal}
  {arXiv:1912.01939}\ } (\bibinfo {year} {2019})}\BibitemShut {NoStop}%
\bibitem [{\citenamefont {Ahmadi}\ \emph {et~al.}(2019)\citenamefont {Ahmadi},
  \citenamefont {Salimi},\ and\ \citenamefont {Khorashad}}]{Ahmadi-refined}%
  \BibitemOpen
  \bibfield  {author} {\bibinfo {author} {\bibfnamefont {B.}~\bibnamefont
  {Ahmadi}}, \bibinfo {author} {\bibfnamefont {S.}~\bibnamefont {Salimi}},\
  and\ \bibinfo {author} {\bibfnamefont {A.~S.}\ \bibnamefont {Khorashad}},\
  }\bibfield  {title} {\bibinfo {title} {Refined definitions of heat and work
  in quantum thermodynamics},\ }\href {https://arxiv.org/abs/1912.01983}
  {\bibfield  {journal} {\bibinfo  {journal} {arXiv:1912.01983}\ } (\bibinfo
  {year} {2019})}\BibitemShut {NoStop}%
\bibitem [{\citenamefont {\ifmmode~\check{S}\else \v{S}\fi{}afr\'anek}\ \emph
  {et~al.}(2019)\citenamefont {\ifmmode~\check{S}\else \v{S}\fi{}afr\'anek},
  \citenamefont {Deutsch},\ and\ \citenamefont {Aguirre}}]{safranek}%
  \BibitemOpen
  \bibfield  {author} {\bibinfo {author} {\bibfnamefont {D.}~\bibnamefont
  {\ifmmode~\check{S}\else \v{S}\fi{}afr\'anek}}, \bibinfo {author}
  {\bibfnamefont {J.~M.}\ \bibnamefont {Deutsch}},\ and\ \bibinfo {author}
  {\bibfnamefont {A.}~\bibnamefont {Aguirre}},\ }\bibfield  {title} {\bibinfo
  {title} {Quantum coarse-grained entropy and thermodynamics},\ }\href
  {https://doi.org/10.1103/PhysRevA.99.010101} {\bibfield  {journal} {\bibinfo
  {journal} {Phys. Rev. A}\ }\textbf {\bibinfo {volume} {99}},\ \bibinfo
  {pages} {010101} (\bibinfo {year} {2019})}\BibitemShut {NoStop}%
\bibitem [{\citenamefont {Sampaio}\ \emph {et~al.}(2018)\citenamefont
  {Sampaio}, \citenamefont {Suomela}, \citenamefont {Ala-Nissila},
  \citenamefont {Anders},\ and\ \citenamefont {Philbin}}]{Sampaio2018}%
  \BibitemOpen
  \bibfield  {author} {\bibinfo {author} {\bibfnamefont {R.}~\bibnamefont
  {Sampaio}}, \bibinfo {author} {\bibfnamefont {S.}~\bibnamefont {Suomela}},
  \bibinfo {author} {\bibfnamefont {T.}~\bibnamefont {Ala-Nissila}}, \bibinfo
  {author} {\bibfnamefont {J.}~\bibnamefont {Anders}},\ and\ \bibinfo {author}
  {\bibfnamefont {T.~G.}\ \bibnamefont {Philbin}},\ }\bibfield  {title}
  {\bibinfo {title} {Quantum work in the {Bohmian} framework},\ }\href
  {https://doi.org/10.1103/PhysRevA.97.012131} {\bibfield  {journal} {\bibinfo
  {journal} {Phys. Rev. A}\ }\textbf {\bibinfo {volume} {97}},\ \bibinfo
  {pages} {012131} (\bibinfo {year} {2018})}\BibitemShut {NoStop}%
\bibitem [{\citenamefont {Mehboudi}\ \emph {et~al.}(2019)\citenamefont
  {Mehboudi}, \citenamefont {Sanpera},\ and\ \citenamefont
  {Correa}}]{mehboudi2019thermometry}%
  \BibitemOpen
  \bibfield  {author} {\bibinfo {author} {\bibfnamefont {M.}~\bibnamefont
  {Mehboudi}}, \bibinfo {author} {\bibfnamefont {A.}~\bibnamefont {Sanpera}},\
  and\ \bibinfo {author} {\bibfnamefont {L.~A.}\ \bibnamefont {Correa}},\
  }\bibfield  {title} {\bibinfo {title} {Thermometry in the quantum regime:
  recent theoretical progress},\ }\href
  {https://doi.org/10.1088/1751-8121/ab2828} {\bibfield  {journal} {\bibinfo
  {journal} {J. Phys. A: Math. Theor.}\ }\textbf {\bibinfo {volume} {52}},\
  \bibinfo {pages} {303001} (\bibinfo {year} {2019})}\BibitemShut {NoStop}%
\bibitem [{\citenamefont {Hovhannisyan}\ and\ \citenamefont
  {Correa}(2018)}]{Correa-manybody-thermometry}%
  \BibitemOpen
  \bibfield  {author} {\bibinfo {author} {\bibfnamefont {K.~V.}\ \bibnamefont
  {Hovhannisyan}}\ and\ \bibinfo {author} {\bibfnamefont {L.~A.}\ \bibnamefont
  {Correa}},\ }\bibfield  {title} {\bibinfo {title} {Measuring the temperature
  of cold many-body quantum systems},\ }\href
  {https://doi.org/10.1103/PhysRevB.98.045101} {\bibfield  {journal} {\bibinfo
  {journal} {Phys. Rev. B}\ }\textbf {\bibinfo {volume} {98}},\ \bibinfo
  {pages} {045101} (\bibinfo {year} {2018})}\BibitemShut {NoStop}%
\bibitem [{\citenamefont {Taranto}\ \emph {et~al.}(2020)\citenamefont
  {Taranto}, \citenamefont {Bakhshinezhad}, \citenamefont {Sch{\"u}ttelkopf},
  \citenamefont {Clivaz},\ and\ \citenamefont {Huber}}]{Bakhshinezhad-cooling}%
  \BibitemOpen
  \bibfield  {author} {\bibinfo {author} {\bibfnamefont {P.}~\bibnamefont
  {Taranto}}, \bibinfo {author} {\bibfnamefont {F.}~\bibnamefont
  {Bakhshinezhad}}, \bibinfo {author} {\bibfnamefont {P.}~\bibnamefont
  {Sch{\"u}ttelkopf}}, \bibinfo {author} {\bibfnamefont {F.}~\bibnamefont
  {Clivaz}},\ and\ \bibinfo {author} {\bibfnamefont {M.}~\bibnamefont
  {Huber}},\ }\bibfield  {title} {\bibinfo {title} {Exponential improvement for
  quantum cooling through finite-memory effects},\ }\href
  {https://doi.org/10.1103/PhysRevApplied.14.054005} {\bibfield  {journal}
  {\bibinfo  {journal} {Phys. Rev. Applied}\ }\textbf {\bibinfo {volume}
  {14}},\ \bibinfo {pages} {054005} (\bibinfo {year} {2020})}\BibitemShut
  {NoStop}%
\bibitem [{\citenamefont {Cotler}\ \emph {et~al.}(2019)\citenamefont {Cotler},
  \citenamefont {Choi}, \citenamefont {Lukin}, \citenamefont {Gharibyan},
  \citenamefont {Grover}, \citenamefont {Tai}, \citenamefont {Rispoli},
  \citenamefont {Schittko}, \citenamefont {Preiss}, \citenamefont {Kaufman},
  \citenamefont {Greiner}, \citenamefont {Pichler},\ and\ \citenamefont
  {Hayden}}]{PRXCotler-virtual-cooling}%
  \BibitemOpen
  \bibfield  {author} {\bibinfo {author} {\bibfnamefont {J.}~\bibnamefont
  {Cotler}}, \bibinfo {author} {\bibfnamefont {S.}~\bibnamefont {Choi}},
  \bibinfo {author} {\bibfnamefont {A.}~\bibnamefont {Lukin}}, \bibinfo
  {author} {\bibfnamefont {H.}~\bibnamefont {Gharibyan}}, \bibinfo {author}
  {\bibfnamefont {T.}~\bibnamefont {Grover}}, \bibinfo {author} {\bibfnamefont
  {M.~E.}\ \bibnamefont {Tai}}, \bibinfo {author} {\bibfnamefont
  {M.}~\bibnamefont {Rispoli}}, \bibinfo {author} {\bibfnamefont
  {R.}~\bibnamefont {Schittko}}, \bibinfo {author} {\bibfnamefont {P.~M.}\
  \bibnamefont {Preiss}}, \bibinfo {author} {\bibfnamefont {A.~M.}\
  \bibnamefont {Kaufman}}, \bibinfo {author} {\bibfnamefont {M.}~\bibnamefont
  {Greiner}}, \bibinfo {author} {\bibfnamefont {H.}~\bibnamefont {Pichler}},\
  and\ \bibinfo {author} {\bibfnamefont {P.}~\bibnamefont {Hayden}},\
  }\bibfield  {title} {\bibinfo {title} {{Quantum Virtual Cooling}},\ }\href
  {https://doi.org/10.1103/PhysRevX.9.031013} {\bibfield  {journal} {\bibinfo
  {journal} {Phys. Rev. X}\ }\textbf {\bibinfo {volume} {9}},\ \bibinfo {pages}
  {031013} (\bibinfo {year} {2019})}\BibitemShut {NoStop}%
\bibitem [{\citenamefont {Raeisi}\ \emph {et~al.}(2019)\citenamefont {Raeisi},
  \citenamefont {Kieferov\'a},\ and\ \citenamefont
  {Mosca}}]{Raeisi-algorCooling}%
  \BibitemOpen
  \bibfield  {author} {\bibinfo {author} {\bibfnamefont {S.}~\bibnamefont
  {Raeisi}}, \bibinfo {author} {\bibfnamefont {M.}~\bibnamefont
  {Kieferov\'a}},\ and\ \bibinfo {author} {\bibfnamefont {M.}~\bibnamefont
  {Mosca}},\ }\bibfield  {title} {\bibinfo {title} {{Novel Technique for Robust
  Optimal Algorithmic Cooling}},\ }\href
  {https://doi.org/10.1103/PhysRevLett.122.220501} {\bibfield  {journal}
  {\bibinfo  {journal} {Phys. Rev. Lett.}\ }\textbf {\bibinfo {volume} {122}},\
  \bibinfo {pages} {220501} (\bibinfo {year} {2019})}\BibitemShut {NoStop}%
\bibitem [{\citenamefont {Alipour}\ \emph
  {et~al.}(2020{\natexlab{b}})\citenamefont {Alipour}, \citenamefont
  {Rezakhani}, \citenamefont {Babu}, \citenamefont {M{\o}lmer}, \citenamefont
  {M{\"o}tt{\"o}nen},\ and\ \citenamefont {Ala-Nissila}}]{ULL}%
  \BibitemOpen
  \bibfield  {author} {\bibinfo {author} {\bibfnamefont {S.}~\bibnamefont
  {Alipour}}, \bibinfo {author} {\bibfnamefont {A.~T.}\ \bibnamefont
  {Rezakhani}}, \bibinfo {author} {\bibfnamefont {A.~P.}\ \bibnamefont {Babu}},
  \bibinfo {author} {\bibfnamefont {K.}~\bibnamefont {M{\o}lmer}}, \bibinfo
  {author} {\bibfnamefont {M.}~\bibnamefont {M{\"o}tt{\"o}nen}},\ and\ \bibinfo
  {author} {\bibfnamefont {T.}~\bibnamefont {Ala-Nissila}},\ }\bibfield
  {title} {\bibinfo {title} {{Correlation-Picture Approach to
  Open-Quantum-System Dynamics}},\ }\href
  {https://doi.org/10.1103/PhysRevX.10.041024} {\bibfield  {journal} {\bibinfo
  {journal} {Phys. Rev. X}\ }\textbf {\bibinfo {volume} {10}},\ \bibinfo
  {pages} {041024} (\bibinfo {year} {2020}{\natexlab{b}})}\BibitemShut
  {NoStop}%
\bibitem [{\citenamefont {Babu}\ \emph {et~al.}(2021)\citenamefont {Babu},
  \citenamefont {Alipour}, \citenamefont {Rezakhani},\ and\ \citenamefont
  {Ala-Nissila}}]{ULL-2}%
  \BibitemOpen
  \bibfield  {author} {\bibinfo {author} {\bibfnamefont {A.~P.}\ \bibnamefont
  {Babu}}, \bibinfo {author} {\bibfnamefont {S.}~\bibnamefont {Alipour}},
  \bibinfo {author} {\bibfnamefont {A.~T.}\ \bibnamefont {Rezakhani}},\ and\
  \bibinfo {author} {\bibfnamefont {T.}~\bibnamefont {Ala-Nissila}},\
  }\bibfield  {title} {\bibinfo {title} {Unfolding correlation from
  open-quantum-system master equations},\ }\href
  {https://arxiv.org/abs/2104.04248} {\bibfield  {journal} {\bibinfo  {journal}
  {arXiv:2104.04248}\ } (\bibinfo {year} {2021})}\BibitemShut {NoStop}%
\bibitem [{\citenamefont {Strasberg}\ \emph
  {et~al.}(2016{\natexlab{a}})\citenamefont {Strasberg}, \citenamefont
  {Schaller}, \citenamefont {Lambert},\ and\ \citenamefont
  {Brandes}}]{strasberg2016nonequilibrium}%
  \BibitemOpen
  \bibfield  {author} {\bibinfo {author} {\bibfnamefont {P.}~\bibnamefont
  {Strasberg}}, \bibinfo {author} {\bibfnamefont {G.}~\bibnamefont {Schaller}},
  \bibinfo {author} {\bibfnamefont {N.}~\bibnamefont {Lambert}},\ and\ \bibinfo
  {author} {\bibfnamefont {T.}~\bibnamefont {Brandes}},\ }\bibfield  {title}
  {\bibinfo {title} {Nonequilibrium thermodynamics in the strong coupling and
  non-markovian regime based on a reaction coordinate mapping},\ }\href
  {https://doi.org/10.1088/1367-2630/18/7/073007} {\bibfield  {journal}
  {\bibinfo  {journal} {New J. Phys.}\ }\textbf {\bibinfo {volume} {18}},\
  \bibinfo {pages} {073007} (\bibinfo {year} {2016}{\natexlab{a}})}\BibitemShut
  {NoStop}%
\bibitem [{\citenamefont {Alipour}\ \emph
  {et~al.}(2020{\natexlab{c}})\citenamefont {Alipour}, \citenamefont {Chenu},
  \citenamefont {Rezakhani},\ and\ \citenamefont {del
  Campo}}]{alipour2020shortcuts}%
  \BibitemOpen
  \bibfield  {author} {\bibinfo {author} {\bibfnamefont {S.}~\bibnamefont
  {Alipour}}, \bibinfo {author} {\bibfnamefont {A.}~\bibnamefont {Chenu}},
  \bibinfo {author} {\bibfnamefont {A.~T.}\ \bibnamefont {Rezakhani}},\ and\
  \bibinfo {author} {\bibfnamefont {A.}~\bibnamefont {del Campo}},\ }\bibfield
  {title} {\bibinfo {title} {{Shortcuts to adiabaticity in driven open quantum
  systems: Balanced gain and loss and non-Markovian evolution}},\ }\href
  {https://doi.org/10.22331/q-2020-09-28-336} {\bibfield  {journal} {\bibinfo
  {journal} {Quantum}\ }\textbf {\bibinfo {volume} {4}},\ \bibinfo {pages}
  {336} (\bibinfo {year} {2020}{\natexlab{c}})}\BibitemShut {NoStop}%
\bibitem [{\citenamefont
  {Seifert}(2016)}]{Seifert-FirstSecondLawStrongCoupling}%
  \BibitemOpen
  \bibfield  {author} {\bibinfo {author} {\bibfnamefont {U.}~\bibnamefont
  {Seifert}},\ }\bibfield  {title} {\bibinfo {title} {{First and Second Law of
  Thermodynamics at Strong Coupling}},\ }\href
  {https://doi.org/10.1103/PhysRevLett.116.020601} {\bibfield  {journal}
  {\bibinfo  {journal} {Phys. Rev. Lett.}\ }\textbf {\bibinfo {volume} {116}},\
  \bibinfo {pages} {020601} (\bibinfo {year} {2016})}\BibitemShut {NoStop}%
\bibitem [{\citenamefont {Jarzynski}(2017)}]{Jarzynski-StronglyCoupledThermo}%
  \BibitemOpen
  \bibfield  {author} {\bibinfo {author} {\bibfnamefont {C.}~\bibnamefont
  {Jarzynski}},\ }\bibfield  {title} {\bibinfo {title} {{Stochastic and
  Macroscopic Thermodynamics of Strongly Coupled Systems}},\ }\href
  {https://doi.org/10.1103/PhysRevX.7.011008} {\bibfield  {journal} {\bibinfo
  {journal} {Phys. Rev. X}\ }\textbf {\bibinfo {volume} {7}},\ \bibinfo {pages}
  {011008} (\bibinfo {year} {2017})}\BibitemShut {NoStop}%
\bibitem [{\citenamefont {Hsiang}\ and\ \citenamefont
  {Hu}(2018)}]{hsiang-StrongCoupling-OperatorThermodynamics}%
  \BibitemOpen
  \bibfield  {author} {\bibinfo {author} {\bibfnamefont {J.-T.}\ \bibnamefont
  {Hsiang}}\ and\ \bibinfo {author} {\bibfnamefont {B.-L.}\ \bibnamefont
  {Hu}},\ }\bibfield  {title} {\bibinfo {title} {Quantum thermodynamics at
  strong coupling: operator thermodynamic functions and relations},\ }\href
  {https://doi.org/10.3390/e20060423} {\bibfield  {journal} {\bibinfo
  {journal} {Entropy}\ }\textbf {\bibinfo {volume} {20}},\ \bibinfo {pages}
  {423} (\bibinfo {year} {2018})}\BibitemShut {NoStop}%
\bibitem [{\citenamefont {Perarnau-Llobet}\ \emph {et~al.}(2018)\citenamefont
  {Perarnau-Llobet}, \citenamefont {Wilming}, \citenamefont {Riera},
  \citenamefont {Gallego},\ and\ \citenamefont
  {Eisert}}]{Perarnau-Llobet-StrongCouplingThermo}%
  \BibitemOpen
  \bibfield  {author} {\bibinfo {author} {\bibfnamefont {M.}~\bibnamefont
  {Perarnau-Llobet}}, \bibinfo {author} {\bibfnamefont {H.}~\bibnamefont
  {Wilming}}, \bibinfo {author} {\bibfnamefont {A.}~\bibnamefont {Riera}},
  \bibinfo {author} {\bibfnamefont {R.}~\bibnamefont {Gallego}},\ and\ \bibinfo
  {author} {\bibfnamefont {J.}~\bibnamefont {Eisert}},\ }\bibfield  {title}
  {\bibinfo {title} {{Strong Coupling Corrections in Quantum Thermodynamics}},\
  }\href {https://doi.org/10.1103/PhysRevLett.120.120602} {\bibfield  {journal}
  {\bibinfo  {journal} {Phys. Rev. Lett.}\ }\textbf {\bibinfo {volume} {120}},\
  \bibinfo {pages} {120602} (\bibinfo {year} {2018})}\BibitemShut {NoStop}%
\bibitem [{\citenamefont {Talkner}\ and\ \citenamefont
  {H\"anggi}(2020)}]{Talkner-StrongCouplingThermo-Classic-Quantum}%
  \BibitemOpen
  \bibfield  {author} {\bibinfo {author} {\bibfnamefont {P.}~\bibnamefont
  {Talkner}}\ and\ \bibinfo {author} {\bibfnamefont {P.}~\bibnamefont
  {H\"anggi}},\ }\bibfield  {title} {\bibinfo {title} {Statistical mechanics
  and thermodynamics at strong coupling: Quantum and classical},\ }\href
  {https://doi.org/10.1103/RevModPhys.92.041002} {\bibfield  {journal}
  {\bibinfo  {journal} {Rev. Mod. Phys.}\ }\textbf {\bibinfo {volume} {92}},\
  \bibinfo {pages} {041002} (\bibinfo {year} {2020})}\BibitemShut {NoStop}%
\bibitem [{\citenamefont {Hossein-Nejad}\ \emph {et~al.}(2015)\citenamefont
  {Hossein-Nejad}, \citenamefont {O'Reilly},\ and\ \citenamefont
  {Olaya-Castro}}]{HosseinNejad-OlayaCastro}%
  \BibitemOpen
  \bibfield  {author} {\bibinfo {author} {\bibfnamefont {H.}~\bibnamefont
  {Hossein-Nejad}}, \bibinfo {author} {\bibfnamefont {E.~J.}\ \bibnamefont
  {O'Reilly}},\ and\ \bibinfo {author} {\bibfnamefont {A.}~\bibnamefont
  {Olaya-Castro}},\ }\bibfield  {title} {\bibinfo {title} {Work, heat and
  entropy production in bipartite quantum systems},\ }\href
  {https://doi.org/10.1088/1367-2630/17/7/075014} {\bibfield  {journal}
  {\bibinfo  {journal} {New J. Phys.}\ }\textbf {\bibinfo {volume} {17}},\
  \bibinfo {pages} {075014} (\bibinfo {year} {2015})}\BibitemShut {NoStop}%
\bibitem [{\citenamefont {Strasberg}\ \emph
  {et~al.}(2016{\natexlab{b}})\citenamefont {Strasberg}, \citenamefont
  {Schaller}, \citenamefont {Lambert},\ and\ \citenamefont
  {Brandes}}]{Strasberg-StrongCoupling-nonMarkovian}%
  \BibitemOpen
  \bibfield  {author} {\bibinfo {author} {\bibfnamefont {P.}~\bibnamefont
  {Strasberg}}, \bibinfo {author} {\bibfnamefont {G.}~\bibnamefont {Schaller}},
  \bibinfo {author} {\bibfnamefont {N.}~\bibnamefont {Lambert}},\ and\ \bibinfo
  {author} {\bibfnamefont {T.}~\bibnamefont {Brandes}},\ }\bibfield  {title}
  {\bibinfo {title} {Nonequilibrium thermodynamics in the strong coupling and
  non-{Markovian} regime based on a reaction coordinate mapping},\ }\href
  {https://doi.org/10.1088/1367-2630/18/7/073007} {\bibfield  {journal}
  {\bibinfo  {journal} {New J. Phys.}\ }\textbf {\bibinfo {volume} {18}},\
  \bibinfo {pages} {073007} (\bibinfo {year} {2016}{\natexlab{b}})}\BibitemShut
  {NoStop}%
\bibitem [{\citenamefont {Iles-Smith}\ \emph {et~al.}(2014)\citenamefont
  {Iles-Smith}, \citenamefont {Lambert},\ and\ \citenamefont {Nazir}}]{nazir}%
  \BibitemOpen
  \bibfield  {author} {\bibinfo {author} {\bibfnamefont {J.}~\bibnamefont
  {Iles-Smith}}, \bibinfo {author} {\bibfnamefont {N.}~\bibnamefont
  {Lambert}},\ and\ \bibinfo {author} {\bibfnamefont {A.}~\bibnamefont
  {Nazir}},\ }\bibfield  {title} {\bibinfo {title} {Environmental dynamics,
  correlations, and the emergence of noncanonical equilibrium states in open
  quantum systems},\ }\href {https://doi.org/10.1103/PhysRevA.90.032114}
  {\bibfield  {journal} {\bibinfo  {journal} {Phys. Rev. A}\ }\textbf {\bibinfo
  {volume} {90}},\ \bibinfo {pages} {032114} (\bibinfo {year}
  {2014})}\BibitemShut {NoStop}%
\bibitem [{\citenamefont {Goyal}\ and\ \citenamefont {Kawai}(2019)}]{kawai}%
  \BibitemOpen
  \bibfield  {author} {\bibinfo {author} {\bibfnamefont {K.}~\bibnamefont
  {Goyal}}\ and\ \bibinfo {author} {\bibfnamefont {R.}~\bibnamefont {Kawai}},\
  }\bibfield  {title} {\bibinfo {title} {Steady state thermodynamics of two
  qubits strongly coupled to bosonic environments},\ }\href
  {https://doi.org/10.1103/PhysRevResearch.1.033018} {\bibfield  {journal}
  {\bibinfo  {journal} {Phys. Rev. Research}\ }\textbf {\bibinfo {volume}
  {1}},\ \bibinfo {pages} {033018} (\bibinfo {year} {2019})}\BibitemShut
  {NoStop}%
\bibitem [{\citenamefont {Ferraro}\ \emph {et~al.}(2012)\citenamefont
  {Ferraro}, \citenamefont {Garc\'{i}a-Saez},\ and\ \citenamefont
  {Ac\'{i}n}}]{Acin-intensiveT}%
  \BibitemOpen
  \bibfield  {author} {\bibinfo {author} {\bibfnamefont {A.}~\bibnamefont
  {Ferraro}}, \bibinfo {author} {\bibfnamefont {A.}~\bibnamefont
  {Garc\'{i}a-Saez}},\ and\ \bibinfo {author} {\bibfnamefont {A.}~\bibnamefont
  {Ac\'{i}n}},\ }\bibfield  {title} {\bibinfo {title} {Intensive temperature
  and quantum correlations for refined quantum measurements},\ }\href
  {https://doi.org/10.1209/0295-5075/98/10009} {\bibfield  {journal} {\bibinfo
  {journal} {Europhys. Lett.}\ }\textbf {\bibinfo {volume} {98}},\ \bibinfo
  {pages} {10009} (\bibinfo {year} {2012})}\BibitemShut {NoStop}%
\bibitem [{\citenamefont {Kliesch}\ \emph {et~al.}(2014)\citenamefont
  {Kliesch}, \citenamefont {Gogolin}, \citenamefont {Kastoryano}, \citenamefont
  {Riera},\ and\ \citenamefont {Eisert}}]{Gogolin-LocalityTemp}%
  \BibitemOpen
  \bibfield  {author} {\bibinfo {author} {\bibfnamefont {M.}~\bibnamefont
  {Kliesch}}, \bibinfo {author} {\bibfnamefont {C.}~\bibnamefont {Gogolin}},
  \bibinfo {author} {\bibfnamefont {M.~J.}\ \bibnamefont {Kastoryano}},
  \bibinfo {author} {\bibfnamefont {A.}~\bibnamefont {Riera}},\ and\ \bibinfo
  {author} {\bibfnamefont {J.}~\bibnamefont {Eisert}},\ }\bibfield  {title}
  {\bibinfo {title} {{Locality of Temperature}},\ }\href
  {https://doi.org/10.1103/PhysRevX.4.031019} {\bibfield  {journal} {\bibinfo
  {journal} {Phys. Rev. X}\ }\textbf {\bibinfo {volume} {4}},\ \bibinfo {pages}
  {031019} (\bibinfo {year} {2014})}\BibitemShut {NoStop}%
\bibitem [{\citenamefont {Hern{\'a}ndez-Santana}\ \emph
  {et~al.}(2015)\citenamefont {Hern{\'a}ndez-Santana}, \citenamefont {Riera},
  \citenamefont {Hovhannisyan}, \citenamefont {Perarnau-Llobet}, \citenamefont
  {Tagliacozzo},\ and\ \citenamefont {Ac{\'\i}n}}]{LocalityTempSpinChain}%
  \BibitemOpen
  \bibfield  {author} {\bibinfo {author} {\bibfnamefont {S.}~\bibnamefont
  {Hern{\'a}ndez-Santana}}, \bibinfo {author} {\bibfnamefont {A.}~\bibnamefont
  {Riera}}, \bibinfo {author} {\bibfnamefont {K.~V.}\ \bibnamefont
  {Hovhannisyan}}, \bibinfo {author} {\bibfnamefont {M.}~\bibnamefont
  {Perarnau-Llobet}}, \bibinfo {author} {\bibfnamefont {L.}~\bibnamefont
  {Tagliacozzo}},\ and\ \bibinfo {author} {\bibfnamefont {A.}~\bibnamefont
  {Ac{\'\i}n}},\ }\bibfield  {title} {\bibinfo {title} {Locality of temperature
  in spin chains},\ }\href {https://doi.org/10.1088/1367-2630/17/8/085007}
  {\bibfield  {journal} {\bibinfo  {journal} {New J. Phys.}\ }\textbf {\bibinfo
  {volume} {17}},\ \bibinfo {pages} {085007} (\bibinfo {year}
  {2015})}\BibitemShut {NoStop}%
\bibitem [{\citenamefont {Pourjafarabadi}\ \emph {et~al.}(2021)\citenamefont
  {Pourjafarabadi}, \citenamefont {Najafzadeh}, \citenamefont {Vaezi},\ and\
  \citenamefont {Vaezi}}]{Vaezi-localT}%
  \BibitemOpen
  \bibfield  {author} {\bibinfo {author} {\bibfnamefont {M.}~\bibnamefont
  {Pourjafarabadi}}, \bibinfo {author} {\bibfnamefont {H.}~\bibnamefont
  {Najafzadeh}}, \bibinfo {author} {\bibfnamefont {M.-S.}\ \bibnamefont
  {Vaezi}},\ and\ \bibinfo {author} {\bibfnamefont {A.}~\bibnamefont {Vaezi}},\
  }\bibfield  {title} {\bibinfo {title} {Entanglement hamiltonian of
  interacting systems: Local temperature approximation and beyond},\ }\href
  {https://doi.org/10.1103/PhysRevResearch.3.013217} {\bibfield  {journal}
  {\bibinfo  {journal} {Phys. Rev. Research}\ }\textbf {\bibinfo {volume}
  {3}},\ \bibinfo {pages} {013217} (\bibinfo {year} {2021})}\BibitemShut
  {NoStop}%
\bibitem [{\citenamefont {Cugliandolo}\ \emph {et~al.}(1997)\citenamefont
  {Cugliandolo}, \citenamefont {Kurchan},\ and\ \citenamefont
  {Peliti}}]{Kurchan}%
  \BibitemOpen
  \bibfield  {author} {\bibinfo {author} {\bibfnamefont {L.~F.}\ \bibnamefont
  {Cugliandolo}}, \bibinfo {author} {\bibfnamefont {J.}~\bibnamefont
  {Kurchan}},\ and\ \bibinfo {author} {\bibfnamefont {L.}~\bibnamefont
  {Peliti}},\ }\bibfield  {title} {\bibinfo {title} {Energy flow, partial
  equilibration, and effective temperatures in systems with slow dynamics},\
  }\href {https://doi.org/10.1103/PhysRevE.55.3898} {\bibfield  {journal}
  {\bibinfo  {journal} {Phys. Rev. E}\ }\textbf {\bibinfo {volume} {55}},\
  \bibinfo {pages} {3898} (\bibinfo {year} {1997})}\BibitemShut {NoStop}%
\bibitem [{\citenamefont {Puglisi}\ \emph {et~al.}(2017)\citenamefont
  {Puglisi}, \citenamefont {Sarracino},\ and\ \citenamefont
  {Vulpiani}}]{puglisi2017temperature}%
  \BibitemOpen
  \bibfield  {author} {\bibinfo {author} {\bibfnamefont {A.}~\bibnamefont
  {Puglisi}}, \bibinfo {author} {\bibfnamefont {A.}~\bibnamefont {Sarracino}},\
  and\ \bibinfo {author} {\bibfnamefont {A.}~\bibnamefont {Vulpiani}},\
  }\bibfield  {title} {\bibinfo {title} {Temperature in and out of equilibrium:
  A review of concepts, tools and attempts},\ }\href
  {https://doi.org/10.1016/j.physrep.2017.09.001} {\bibfield  {journal}
  {\bibinfo  {journal} {Phys. Rep.}\ }\textbf {\bibinfo {volume} {709}},\
  \bibinfo {pages} {1} (\bibinfo {year} {2017})}\BibitemShut {NoStop}%
\bibitem [{\citenamefont {Martens}\ \emph {et~al.}(2009)\citenamefont
  {Martens}, \citenamefont {Bertin},\ and\ \citenamefont
  {Droz}}]{Martens-FlucDissipTemp}%
  \BibitemOpen
  \bibfield  {author} {\bibinfo {author} {\bibfnamefont {K.}~\bibnamefont
  {Martens}}, \bibinfo {author} {\bibfnamefont {E.}~\bibnamefont {Bertin}},\
  and\ \bibinfo {author} {\bibfnamefont {M.}~\bibnamefont {Droz}},\ }\bibfield
  {title} {\bibinfo {title} {{Dependence of the Fluctuation-Dissipation
  Temperature on the Choice of Observable}},\ }\href
  {https://doi.org/10.1103/PhysRevLett.103.260602} {\bibfield  {journal}
  {\bibinfo  {journal} {Phys. Rev. Lett.}\ }\textbf {\bibinfo {volume} {103}},\
  \bibinfo {pages} {260602} (\bibinfo {year} {2009})}\BibitemShut {NoStop}%
\bibitem [{\citenamefont {Callen}(1985)}]{book:Callen}%
  \BibitemOpen
  \bibfield  {author} {\bibinfo {author} {\bibfnamefont {H.~B.}\ \bibnamefont
  {Callen}},\ }\href@noop {} {\emph {\bibinfo {title} {Thermodynamics and
  Introduction to Thermostatistics}}}\ (\bibinfo  {publisher} {John Wiley \&
  Sons},\ \bibinfo {address} {New York},\ \bibinfo {year} {1985})\BibitemShut
  {NoStop}%
\bibitem [{\citenamefont {Vallejo}\ \emph {et~al.}(2020)\citenamefont
  {Vallejo}, \citenamefont {Romanelli},\ and\ \citenamefont
  {Don{\'a}ngelo}}]{vallejo2020temperature}%
  \BibitemOpen
  \bibfield  {author} {\bibinfo {author} {\bibfnamefont {A.}~\bibnamefont
  {Vallejo}}, \bibinfo {author} {\bibfnamefont {A.}~\bibnamefont {Romanelli}},\
  and\ \bibinfo {author} {\bibfnamefont {R.}~\bibnamefont {Don{\'a}ngelo}},\
  }\bibfield  {title} {\bibinfo {title} {Temperature of a finite-dimensional
  quantum system},\ }\href {https://arxiv.org/pdf/2005.00261.pdf} {\bibfield
  {journal} {\bibinfo  {journal} {arXiv:2005.00261}\ } (\bibinfo {year}
  {2020})}\BibitemShut {NoStop}%
\bibitem [{\citenamefont {Fick}\ and\ \citenamefont
  {Sauermann}(1986)}]{book:Fick-Sauermann}%
  \BibitemOpen
  \bibfield  {author} {\bibinfo {author} {\bibfnamefont {E.}~\bibnamefont
  {Fick}}\ and\ \bibinfo {author} {\bibfnamefont {G.}~\bibnamefont
  {Sauermann}},\ }\href@noop {} {\emph {\bibinfo {title} {The Quantum
  Statistics of Dynamic Processes}}}\ (\bibinfo  {publisher} {Springer},\
  \bibinfo {address} {Berlin},\ \bibinfo {year} {1986})\BibitemShut {NoStop}%
\bibitem [{\citenamefont {Breuer}\ and\ \citenamefont
  {Petruccione}(2007)}]{BreuerBook}%
  \BibitemOpen
  \bibfield  {author} {\bibinfo {author} {\bibfnamefont {H.-P.}\ \bibnamefont
  {Breuer}}\ and\ \bibinfo {author} {\bibfnamefont {F.}~\bibnamefont
  {Petruccione}},\ }\href@noop {} {\emph {\bibinfo {title} {The Theory of Open
  Quantum Systems}}}\ (\bibinfo  {publisher} {Oxford University Press},\
  \bibinfo {address} {New York},\ \bibinfo {year} {2007})\BibitemShut {NoStop}%
\bibitem [{SM()}]{SM}%
  \BibitemOpen
  \href@noop {} {}\bibinfo {note} {See Supplemental Material at
  \url{http://www}.}\BibitemShut {Stop}%
\bibitem [{\citenamefont {Li}\ and\ \citenamefont {Haldane}(2008)}]{ent-ham}%
  \BibitemOpen
  \bibfield  {author} {\bibinfo {author} {\bibfnamefont {H.}~\bibnamefont
  {Li}}\ and\ \bibinfo {author} {\bibfnamefont {F.~D.~M.}\ \bibnamefont
  {Haldane}},\ }\bibfield  {title} {\bibinfo {title} {{Entanglement Spectrum as
  a Generalization of Entanglement Entropy: Identification of Topological Order
  in Non-Abelian Fractional Quantum Hall Effect States}},\ }\href
  {https://doi.org/10.1103/PhysRevLett.101.010504} {\bibfield  {journal}
  {\bibinfo  {journal} {Phys. Rev. Lett.}\ }\textbf {\bibinfo {volume} {101}},\
  \bibinfo {pages} {010504} (\bibinfo {year} {2008})}\BibitemShut {NoStop}%
\bibitem [{\citenamefont {Schliemann}(2011)}]{ent-spec}%
  \BibitemOpen
  \bibfield  {author} {\bibinfo {author} {\bibfnamefont {J.}~\bibnamefont
  {Schliemann}},\ }\bibfield  {title} {\bibinfo {title} {Entanglement spectrum
  and entanglement thermodynamics of quantum {Hall} bilayers at
  $\ensuremath{\nu}=1$},\ }\href {https://doi.org/10.1103/PhysRevB.83.115322}
  {\bibfield  {journal} {\bibinfo  {journal} {Phys. Rev. B}\ }\textbf {\bibinfo
  {volume} {83}},\ \bibinfo {pages} {115322} (\bibinfo {year}
  {2011})}\BibitemShut {NoStop}%
\bibitem [{\citenamefont {Smorodinsky}(1984)}]{Smorodinsky-temperature}%
  \BibitemOpen
  \bibfield  {author} {\bibinfo {author} {\bibfnamefont {Y.~A.}\ \bibnamefont
  {Smorodinsky}},\ }\href@noop {} {\emph {\bibinfo {title} {Temperature}}}\
  (\bibinfo  {publisher} {Mir},\ \bibinfo {address} {Moscow},\ \bibinfo {year}
  {1984})\BibitemShut {NoStop}%
\bibitem [{\citenamefont {Purcell}\ and\ \citenamefont
  {Pound}(1951)}]{Purcell-NeqTemp}%
  \BibitemOpen
  \bibfield  {author} {\bibinfo {author} {\bibfnamefont {E.~M.}\ \bibnamefont
  {Purcell}}\ and\ \bibinfo {author} {\bibfnamefont {R.~V.}\ \bibnamefont
  {Pound}},\ }\bibfield  {title} {\bibinfo {title} {A nuclear spin system at
  negative temperature},\ }\href {https://doi.org/10.1103/PhysRev.81.279}
  {\bibfield  {journal} {\bibinfo  {journal} {Phys. Rev.}\ }\textbf {\bibinfo
  {volume} {81}},\ \bibinfo {pages} {279} (\bibinfo {year} {1951})}\BibitemShut
  {NoStop}%
\bibitem [{\citenamefont {Oja}\ and\ \citenamefont
  {Lounasmaa}(1997)}]{Oja-NegTemp}%
  \BibitemOpen
  \bibfield  {author} {\bibinfo {author} {\bibfnamefont {A.~S.}\ \bibnamefont
  {Oja}}\ and\ \bibinfo {author} {\bibfnamefont {O.~V.}\ \bibnamefont
  {Lounasmaa}},\ }\bibfield  {title} {\bibinfo {title} {Nuclear magnetic
  ordering in simple metals at positive and negative nanokelvin temperatures},\
  }\href {https://doi.org/10.1103/RevModPhys.69.1} {\bibfield  {journal}
  {\bibinfo  {journal} {Rev. Mod. Phys.}\ }\textbf {\bibinfo {volume} {69}},\
  \bibinfo {pages} {1} (\bibinfo {year} {1997})}\BibitemShut {NoStop}%
\bibitem [{\citenamefont {Medley}\ \emph {et~al.}(2011)\citenamefont {Medley},
  \citenamefont {Weld}, \citenamefont {Miyake}, \citenamefont {Pritchard},\
  and\ \citenamefont {Ketterle}}]{Medley-NegTemp}%
  \BibitemOpen
  \bibfield  {author} {\bibinfo {author} {\bibfnamefont {P.}~\bibnamefont
  {Medley}}, \bibinfo {author} {\bibfnamefont {D.~M.}\ \bibnamefont {Weld}},
  \bibinfo {author} {\bibfnamefont {H.}~\bibnamefont {Miyake}}, \bibinfo
  {author} {\bibfnamefont {D.~E.}\ \bibnamefont {Pritchard}},\ and\ \bibinfo
  {author} {\bibfnamefont {W.}~\bibnamefont {Ketterle}},\ }\bibfield  {title}
  {\bibinfo {title} {{Spin Gradient Demagnetization Cooling of Ultracold
  Atoms}},\ }\href {https://doi.org/10.1103/PhysRevLett.106.195301} {\bibfield
  {journal} {\bibinfo  {journal} {Phys. Rev. Lett.}\ }\textbf {\bibinfo
  {volume} {106}},\ \bibinfo {pages} {195301} (\bibinfo {year}
  {2011})}\BibitemShut {NoStop}%
\bibitem [{\citenamefont {Braun}\ \emph {et~al.}(2013)\citenamefont {Braun},
  \citenamefont {Ronzheimer}, \citenamefont {Schreiber}, \citenamefont
  {Hodgman}, \citenamefont {Rom}, \citenamefont {Bloch},\ and\ \citenamefont
  {Schneider}}]{braun-negativeTemp}%
  \BibitemOpen
  \bibfield  {author} {\bibinfo {author} {\bibfnamefont {S.}~\bibnamefont
  {Braun}}, \bibinfo {author} {\bibfnamefont {J.~P.}\ \bibnamefont
  {Ronzheimer}}, \bibinfo {author} {\bibfnamefont {M.}~\bibnamefont
  {Schreiber}}, \bibinfo {author} {\bibfnamefont {S.~S.}\ \bibnamefont
  {Hodgman}}, \bibinfo {author} {\bibfnamefont {T.}~\bibnamefont {Rom}},
  \bibinfo {author} {\bibfnamefont {I.}~\bibnamefont {Bloch}},\ and\ \bibinfo
  {author} {\bibfnamefont {U.}~\bibnamefont {Schneider}},\ }\bibfield  {title}
  {\bibinfo {title} {Negative absolute temperature for motional degrees of
  freedom},\ }\href {https://doi.org/10.1126/science.1227831} {\bibfield
  {journal} {\bibinfo  {journal} {Science}\ }\textbf {\bibinfo {volume}
  {339}},\ \bibinfo {pages} {52} (\bibinfo {year} {2013})}\BibitemShut
  {NoStop}%
\bibitem [{\citenamefont {Langen}\ \emph {et~al.}(2015)\citenamefont {Langen},
  \citenamefont {Erne}, \citenamefont {Geiger}, \citenamefont {Rauer},
  \citenamefont {Schweigler}, \citenamefont {Kuhnert}, \citenamefont
  {Rohringer}, \citenamefont {Mazets}, \citenamefont {Gasenzer},\ and\
  \citenamefont {Schmiedmayer}}]{Schmiedmayer-GeneralizedGibbsObservation}%
  \BibitemOpen
  \bibfield  {author} {\bibinfo {author} {\bibfnamefont {T.}~\bibnamefont
  {Langen}}, \bibinfo {author} {\bibfnamefont {S.}~\bibnamefont {Erne}},
  \bibinfo {author} {\bibfnamefont {R.}~\bibnamefont {Geiger}}, \bibinfo
  {author} {\bibfnamefont {B.}~\bibnamefont {Rauer}}, \bibinfo {author}
  {\bibfnamefont {T.}~\bibnamefont {Schweigler}}, \bibinfo {author}
  {\bibfnamefont {M.}~\bibnamefont {Kuhnert}}, \bibinfo {author} {\bibfnamefont
  {W.}~\bibnamefont {Rohringer}}, \bibinfo {author} {\bibfnamefont {I.~E.}\
  \bibnamefont {Mazets}}, \bibinfo {author} {\bibfnamefont {T.}~\bibnamefont
  {Gasenzer}},\ and\ \bibinfo {author} {\bibfnamefont {J.}~\bibnamefont
  {Schmiedmayer}},\ }\bibfield  {title} {\bibinfo {title} {Experimental
  observation of a generalized {Gibbs} ensemble},\ }\href
  {https://doi.org/10.1126/science.1257026} {\bibfield  {journal} {\bibinfo
  {journal} {Science}\ }\textbf {\bibinfo {volume} {348}},\ \bibinfo {pages}
  {207} (\bibinfo {year} {2015})}\BibitemShut {NoStop}%
\bibitem [{\citenamefont {Weimer}\ \emph {et~al.}(2008)\citenamefont {Weimer},
  \citenamefont {Henrich}, \citenamefont {Rempp}, \citenamefont
  {Schr{\"o}der},\ and\ \citenamefont {Mahler}}]{Weimer-Mahler-WorkHeat}%
  \BibitemOpen
  \bibfield  {author} {\bibinfo {author} {\bibfnamefont {H.}~\bibnamefont
  {Weimer}}, \bibinfo {author} {\bibfnamefont {M.~J.}\ \bibnamefont {Henrich}},
  \bibinfo {author} {\bibfnamefont {F.}~\bibnamefont {Rempp}}, \bibinfo
  {author} {\bibfnamefont {H.}~\bibnamefont {Schr{\"o}der}},\ and\ \bibinfo
  {author} {\bibfnamefont {G.}~\bibnamefont {Mahler}},\ }\bibfield  {title}
  {\bibinfo {title} {Local effective dynamics of quantum systems: A generalized
  approach to work and heat},\ }\href
  {https://doi.org/10.1209/0295-5075/83/30008} {\bibfield  {journal} {\bibinfo
  {journal} {Europhys. Lett.}\ }\textbf {\bibinfo {volume} {83}},\ \bibinfo
  {pages} {30008} (\bibinfo {year} {2008})}\BibitemShut {NoStop}%
\end{thebibliography}
\end{document}